\documentclass[prx,twocolumn,english,superscriptaddress,longbibliography]{revtex4-2}
\usepackage[T1]{fontenc}
\usepackage[latin9]{inputenc}
\usepackage{graphicx}
\usepackage{amsmath}  
\usepackage{xfrac}
\usepackage{txfonts}
\usepackage[mathscr]{euscript}

\makeatletter
\usepackage{babel}
\usepackage{color}
\usepackage[colorlinks = true,
            linkcolor = blue,
            urlcolor  = blue,
            citecolor = blue,
            anchorcolor = blue]{hyperref}

\makeatother
\begin{document}


%
%

\title{Spin exchange Hamiltonian and topological degeneracies in elemental gadolinium}
	
	\author{A. Scheie}
	\email{scheieao@ornl.gov}
	\affiliation{Neutron Scattering Division, Oak Ridge National Laboratory, Oak Ridge, TN 37831, USA}
	
	\author{Pontus Laurell}
	\affiliation{Computational Sciences and Engineering Division, Oak Ridge National Laboratory, Oak Ridge, TN 37831, USA}
	\address{Department of Physics and Astronomy, University of Tennessee, Knoxville, TN 37996, USA.}
	
	\author{P. A. McClarty}
	\affiliation{Max Planck Institute for the Physics of Complex Systems, N\"othnitzer Str. 38, 01187 Dresden, Germany}
    
	\author{G. E. Granroth}
	\affiliation{Neutron Scattering Division, Oak Ridge National Laboratory, Oak Ridge, TN 37831, USA}

	\author{M. B. Stone}
	\affiliation{Neutron Scattering Division, Oak Ridge National Laboratory, Oak Ridge, TN 37831, USA}
	
	\author{R. Moessner}
	\affiliation{Max Planck Institute for the Physics of Complex Systems, N\"othnitzer Str. 38, 01187 Dresden, Germany}

	\author{S. E. Nagler}
	\affiliation{Neutron Scattering Division, Oak Ridge National Laboratory, Oak Ridge, TN 37831, USA}
	\affiliation{Quantum Science Center, Oak Ridge National Laboratory, Tennessee 37831, USA}
	
	\date{\today}

	\begin{abstract}
    We present a comprehensive study of the magnetic exchange Hamiltonian of elemental Gadolinium. We use neutron scattering to measure the magnon spectrum over the entire Brillouin zone, and fit the excitations to a spin wave model to extract the first 26 nearest neighbor magnetic exchange interactions with rigorously defined uncertainty. We find these exchange interactions to follow RKKY behavior, oscillating from ferromagnetic to antiferromagnetic as a function of distance. Finally, we discuss the topological features and degeneracies in Gd, and HCP ferromagnets in general. We show theoretically how, with asymmetric exchange, topological properties could be tuned with a magnetic field.
	\end{abstract}
	
	\maketitle

\section{Introduction}

Gadolinium (Gd) is one of the few elemental ferromagnets \cite{ Urbain_1935}. It is considered a ``critical material'' to industry because of its unique magnetic properties \cite{chu2011critical,schuler2011study,zhou2020rare}, and yet the mechanisms behind its magnetism are not fully understood. In this study, we use inelastic neutron scattering to determine the magnetic exchange Hamiltonian of elemental Gd to the 26th neighbor exchange, showing the exchange approximately follows an Ruderman-Kittel-Kasuya-Yoshida (RKKY) model. In a separate paper \cite{Scheie_Gd_PRL} we also show that these spin waves yield topological features; in this paper we extend this topology discussion with further details about anisotropic HCP rare earths.

\begin{figure}
	\centering\includegraphics[width=0.3\textwidth]{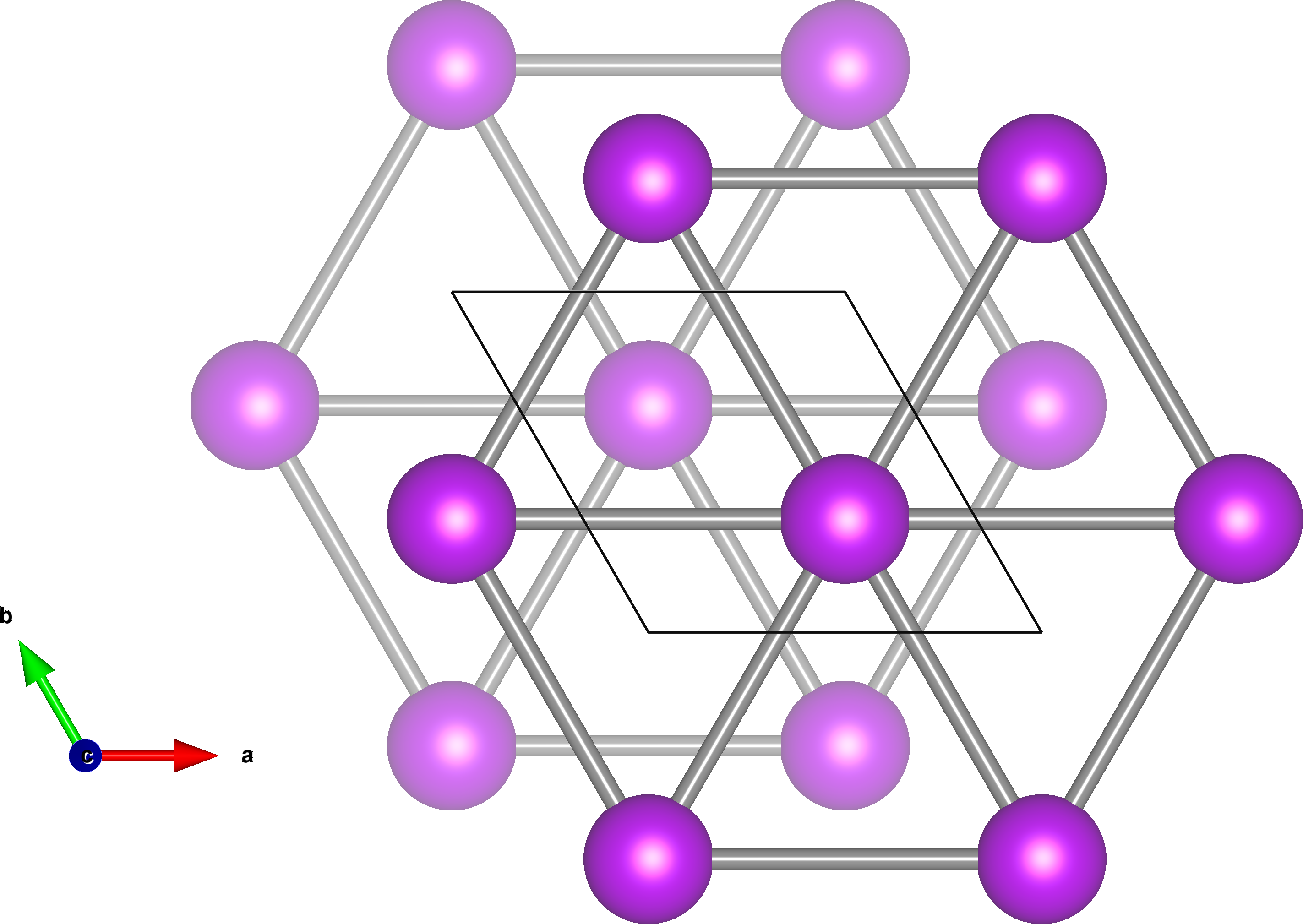}
	\caption{HCP crystal structure of Gd, with layers of triangular lattices.}
	\label{flo:Gd}
\end{figure}

The Gd crystal structure is hexagonal close packed (HCP), shown in Fig. \ref{flo:Gd}.
Elemental Gd orders ferromagnetically \cite{Urbain_1935} with a Curie temperature $T_c = 293$~K \cite{Nigh_1963, Cable_1968}. Its magnetism is almost perfectly isotropic: the first three valence electrons are itinerant, leaving an effective Gd$^{3+}$ at each site \cite{Moon_1972} with a quenched orbital moment and well-defined $S=7/2$ \cite{Kip_1953}. 
Small anisotropies do exist in Gd \cite{Franse_1980}, which vary as a function of temperature \cite{Coey1999,Cable_1968} (leading to a ferromagnetic spin polarization $30^{\circ}$ from $c$ at the lowest temperatures \cite{Cable_1968}) and appear to be from interaction between the itinerant electrons and localized $4f$ electrons \cite{Tosti_2005, PhysRevB.79.054406}. 

The Gd inelastic neutron spectrum was first measured in 1970  \cite{Koehler_1970}, and it is a textbook example of ferromagnetic spin waves \cite{Squires,Jensen+Mackintosh}.  
However the original measurements, performed on a triple-axis spectrometer, were only along high-symmetry directions. This led to limitations in fitting a spin wave Hamiltonian \cite{Lindgard_1978}: in-plane and out-of-plane exchanges could not be fitted simultaneously. Fortunately, neutron spectrometers have dramatically improved over the last 50 years \cite{stone2014comparison}. 
In this study, we measure the inelastic neutron spectrum of elemental Gd over the entire Brillouin zone using modern instrumentation. This allows us to rigorously fit the excitations to a full spin wave model, compare to RKKY exchange strengths, and observe topological features in the data.

This paper is organized as follows: section \ref{sec:Exp} explains the neutron scattering experiment on Gd, section \ref{sec:SpinWaveFit} shows the spin wave fits to the scattering data and compares the exchanges to RKKY, and section \ref{sec:topology} discusses the topology of the Gd magnon band structure.

\section{Experiment} \label{sec:Exp}

We measured the spin waves of Gd using the SEQUOIA spectrometer \cite{Granroth2010,Granroth2006} at the ORNL SNS \cite{mason2006spallation}.
Because most Gd isotopes have an extremely high neutron absorption cross section, we measured a 12~g isotopically enriched $^{160}$Gd single crystal (in fact, the same 99.99\% enriched crystal as was used in Ref. \cite{Koehler_1970}). The sample was mounted with the $(hh\ell)$ plane horizontal in a closed cycle refrigerator, and cooled to 5 K. We measured the neutron spectrum at incident energies $E_i = 50$~meV and 100~meV. We rotated the sample about the vertical axis in one degree steps over 180 degrees. See Appendix~\ref{app:ExpDetails} for further details. Data were reduced and symmetrized \cite{Mantid2014} to fill out the full Brillouin zone. Measurements were performed with the SNS operating at 1.4~MW over the course of two days.

\begin{figure*}
	\centering\includegraphics[width=\textwidth]{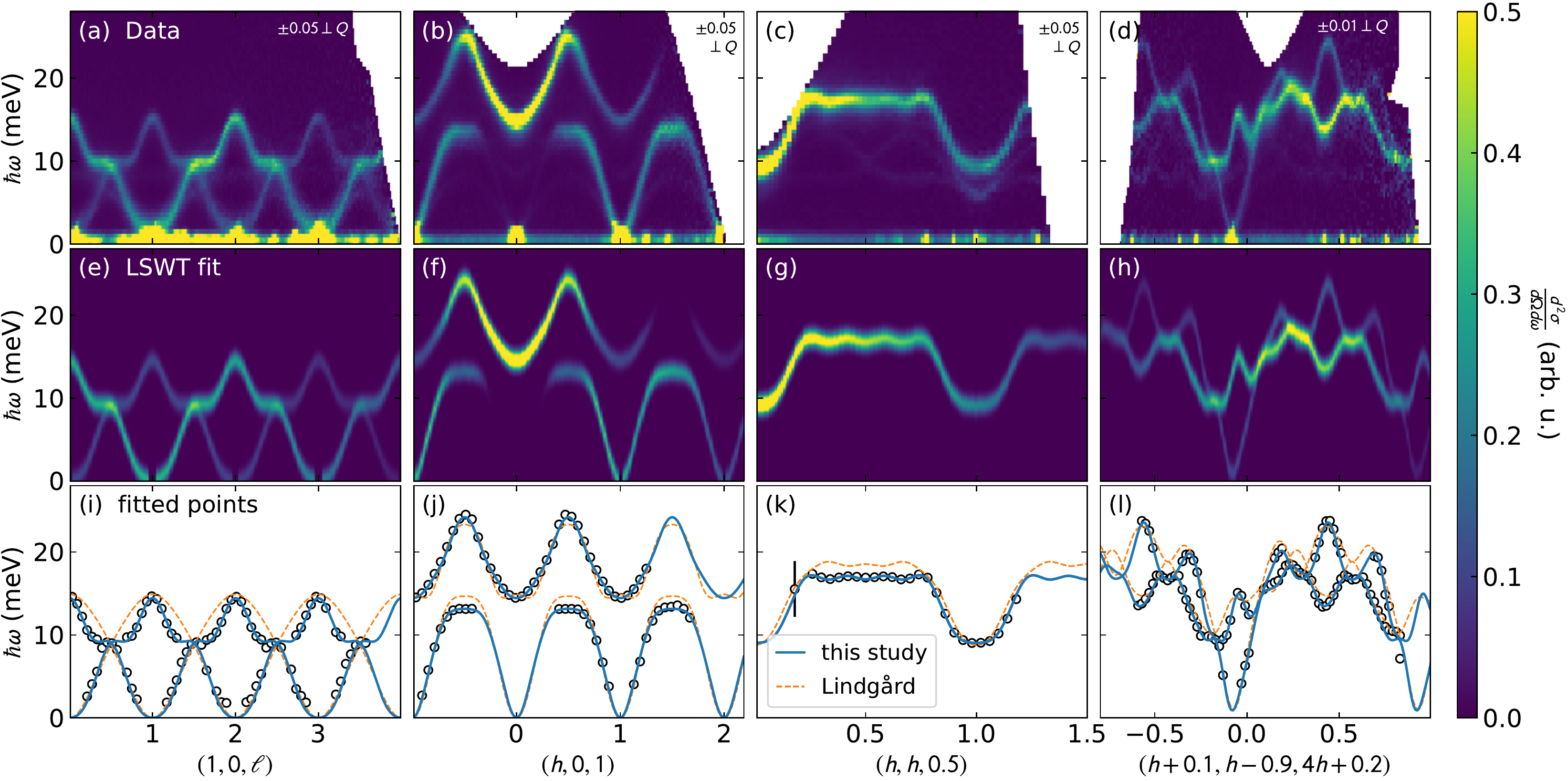}
	\caption{Measured and fitted spin wave spectra of Gd. The top row (a)-(d) shows the measured spin wave spectra of Gd. The $Q$ width in reciprocal lattice units (r.l.u.) in the direction perpendicular to the $x$ axis is shown in the upper right of each panel.  The middle row (e)-(h) shows the LSWT calculated spectrum from the best fit Hamiltonian in Table \ref{tab:FittedJ}. The bottom row (i)-(l) shows a portion of the data points used in the fit (black circles), and the fitted dispersion from this study (blue solid line) and Lindg\aa rd \cite{Lindgard_1978} (orange dashed line).}
	\label{flo:SpinWaves}
\end{figure*}

 Some slices of the Gd scattering data are shown in Fig. \ref{flo:SpinWaves}(a)-(d) with constant $Q$ cuts in Fig. \ref{flo:SpinWavesQ}; see the Supplemental Information \cite{SuppMat} for more plots. The magnon modes are sharp and well defined even at the top of the dispersion, which indicates localized moments as expected. Because of the large $S=7/2$ spin, the magnon intensity is much larger than phonon intensity at 5~K for the measured wavevectors. The phonon modes have similar shaped dispersions to the magnons due to the HCP structure: in certain slices [like $(h,h,0.5)$ in Fig. \ref{flo:SpinWaves}(c)], the phonon modes appear as faint modes at lower energies.  However, the phonons can be distinguished by the fact that their intensity grows with momentum transfer $|Q|$, while the magnon intensity decreases (see Supplemental Information for details \cite{SuppMat}).

\begin{figure}
	\centering\includegraphics[width=0.47\textwidth]{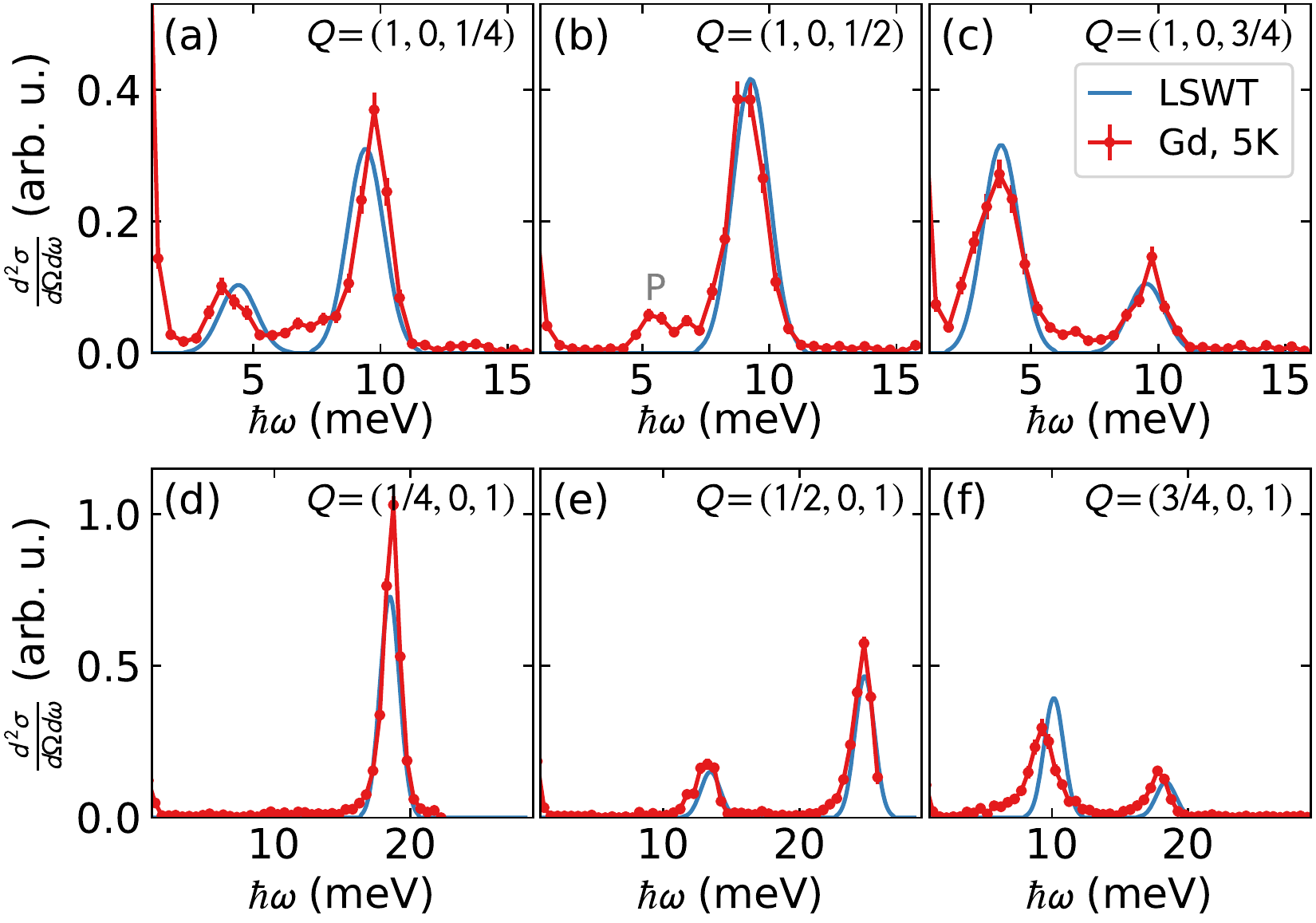}
	\caption{Constant $Q$ cuts of Gd scattering along the $(1,0,\ell)$ direction [top row, from Fig. \ref{flo:SpinWaves}(a)] and $(h,0,1)$ direction [bottom row, from Fig. \ref{flo:SpinWaves}(b)] compared to the best fit LSWT calculation. The magnon peaks are very clearly visible in the data. A phonon is visible in panel (b) at 6~meV, with much smaller intensity than the magnons. LSWT simulations have a energy peak full width half maximum of 1.5 meV, which was chosen to visually match the experimental width at all energies.}
	\label{flo:SpinWavesQ}
\end{figure}

\section{Spin wave fit} \label{sec:SpinWaveFit}

To date, the most comprehensive Hamiltonian fit to Gd spin waves was performed by Lindg\aa rd in 1978 \cite{Lindgard_1978}. This analysis used the inelastic neutron scattering data collected by Koehler in 1970 \cite{Koehler_1970}. Lindg\aa rd included 11 exchange parameters fitted to $hk$ plane scattering, but fitted the $\ell$ direction separately to a $J({\bf Q})$ model (effectively calculating Fourier components) rather than including it in the fit. With our data set, we are able to perform a full, comprehensive Hamiltonian fit to the data.

The magnon dispersion of a HCP Heisenberg ferromagnet described by 
$\mathcal{H} = \sum_{ij} J_{ij} {\bf S}_i \cdot {\bf S}_j$ (here ${\bf S}_i$ are $S=7/2$ operators) is derived in Ref. \cite{Jensen+Mackintosh}. Although the magnon bands appear gapless in our neutron data, magnetic torque measurements reveal the single-ion anisotropy of Gd to be $37 \> {\rm \mu eV}$. To first order, this produces a linear offset of all dispersion bands \cite{Jensen+Mackintosh} of $37 \> {\rm \mu eV}$, and we incorporated this effect into our model. Beyond this single-ion anisotropy, we assumed that all exchanges are isotropic in accord with Gd's quenched orbital moment.


\begin{figure*}
	\includegraphics[width=\linewidth]{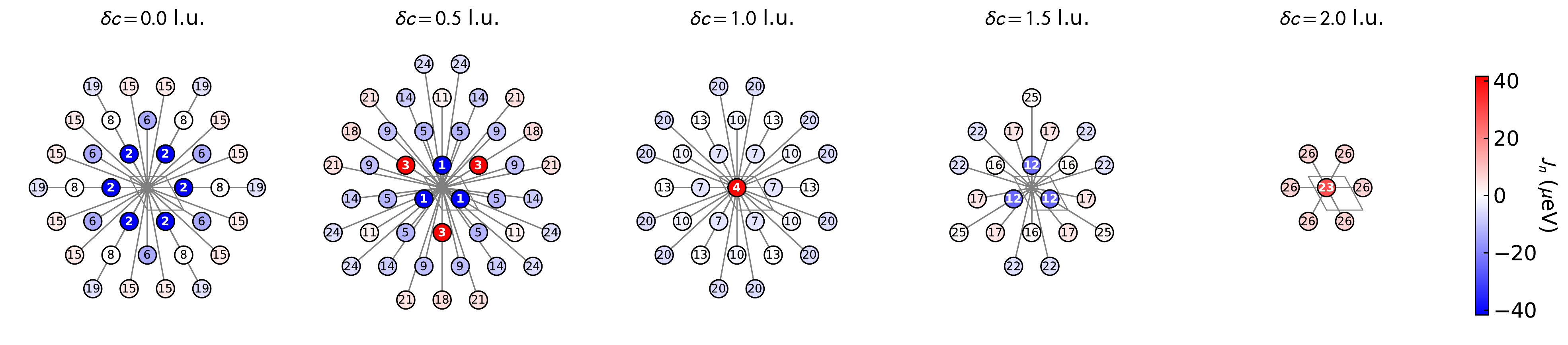}  
	\caption{Neighbor bond length and magnitude of the refined exchange for Gd. The five panels indicate different layers along the $c$ axis, where $\delta c$ in each panel indicates the bond length along the $c$ axis. The number in each circle denotes the neighbor number. The color indicates the fitted exchange interaction: ferromagnetic exchange is indicated by blue, antiferromagnetic by red, with the magnitude indicated by the color bar on the right.}
	\label{flo:Gd_exchanges_color}
\end{figure*}

In order to fit the magnon dispersions to the observed energies, we extracted the magnon dispersion from the data by fitting the magnon modes with Gaussian peaks as a function of  energy transfer for fixed values of wave-vector. We used over 42 different $\bf Q$ vs $\hbar \omega$ slices through our data (see Fig \ref{flo:SpinWaves}(i)-(l) for examples and the Supplemental Information \cite{SuppMat} for a complete list), which yielded a total of 2309 
unique $\bf Q$ points. We found that including data away from high-symmetry directions was important for constraining the fit, hence the large number of slices.
These mode energy points were used to define a global $\chi^2$.

To determine the exchange constant terms, we compared the analytical form of the dispersion \cite{Jensen+Mackintosh} to the determined energy transfer of the 2309 unique wave-vectors including an overall offset of 0.037 meV to account for the anisotropy as described earlier.
The fitting procedure itself involved a stochastic simulated annealing method based on Scipy's minimization package \cite{virtanen2020scipy} and is described in detail in Appendix \ref{app:FittingProcedure}. We found that at least 26 nearest neighbors are required to accurately describe the details of the Gd magnon modes, and all subsequent neighbors had uncertainties overlapping with zero. The best fit parameters with uncertainty are shown in Table \ref{tab:FittedJ}, and are visually depicted in Fig. \ref{flo:Gd_exchanges_color}. The simulated neutron spectrum is plotted in Fig. \ref{flo:SpinWaves}(e)-(h). In every cut, the simulation matches the data quite well.

\begin{table}
	\caption{Refined exchange constants for Gd, in units of $\mu$eV from fine binned data. Positive values indicate antiferromagnetic exchange, negative values indicate ferromagnetic exchange. Error bars indicate one standard deviation.}
	\begin{ruledtabular}
	\begin{tabular}{c c | c c | c c}
$J_{1}=$  &  $ -138 \pm 8$ &  $J_{10}=$  &  $ -2 \pm 2$ &  $J_{19}=$  &  $ -5 \pm 3$ \\
$J_{2}=$  &  $ -174 \pm 4$ &  $J_{11}=$  &  $ 0 \pm 20$ &  $J_{20}=$  &  $ -5.6 \pm 1.3$ \\
$J_{3}=$  &  $ 50 \pm 20$ &  $J_{12}=$  &  $ -25 \pm 6$ &  $J_{21}=$  &  $ 5 \pm 9$ \\
$J_{4}=$  &  $ 41 \pm 6$ &  $J_{13}=$  &  $ 0 \pm 2$ &  $J_{22}=$  &  $ -5 \pm 7$ \\
$J_{5}=$  &  $ -10 \pm 20$ &  $J_{14}=$  &  $ -8 \pm 10$ &  $J_{23}=$  &  $ 29 \pm 7$ \\
$J_{6}=$  &  $ -14 \pm 4$ &  $J_{15}=$  &  $ 4 \pm 2$ &  $J_{24}=$  &  $ -5 \pm 9$ \\
$J_{7}=$  &  $ -4 \pm 2$ &  $J_{16}=$  &  $ 1 \pm 7$ &  $J_{25}=$  &  $ 1 \pm 8$ \\
$J_{8}=$  &  $ 0 \pm 3$ &  $J_{17}=$  &  $ 4 \pm 8$ &  $J_{26}=$  &  $ 7 \pm 2$ \\
$J_{9}=$  &  $ -10 \pm 20$ &  $J_{18}=$  &  $ 10 \pm 20$ &   \\
	\end{tabular}\end{ruledtabular}
	\label{tab:FittedJ}
\end{table}

\begin{figure}
	\centering
	\includegraphics[width=0.98\linewidth]{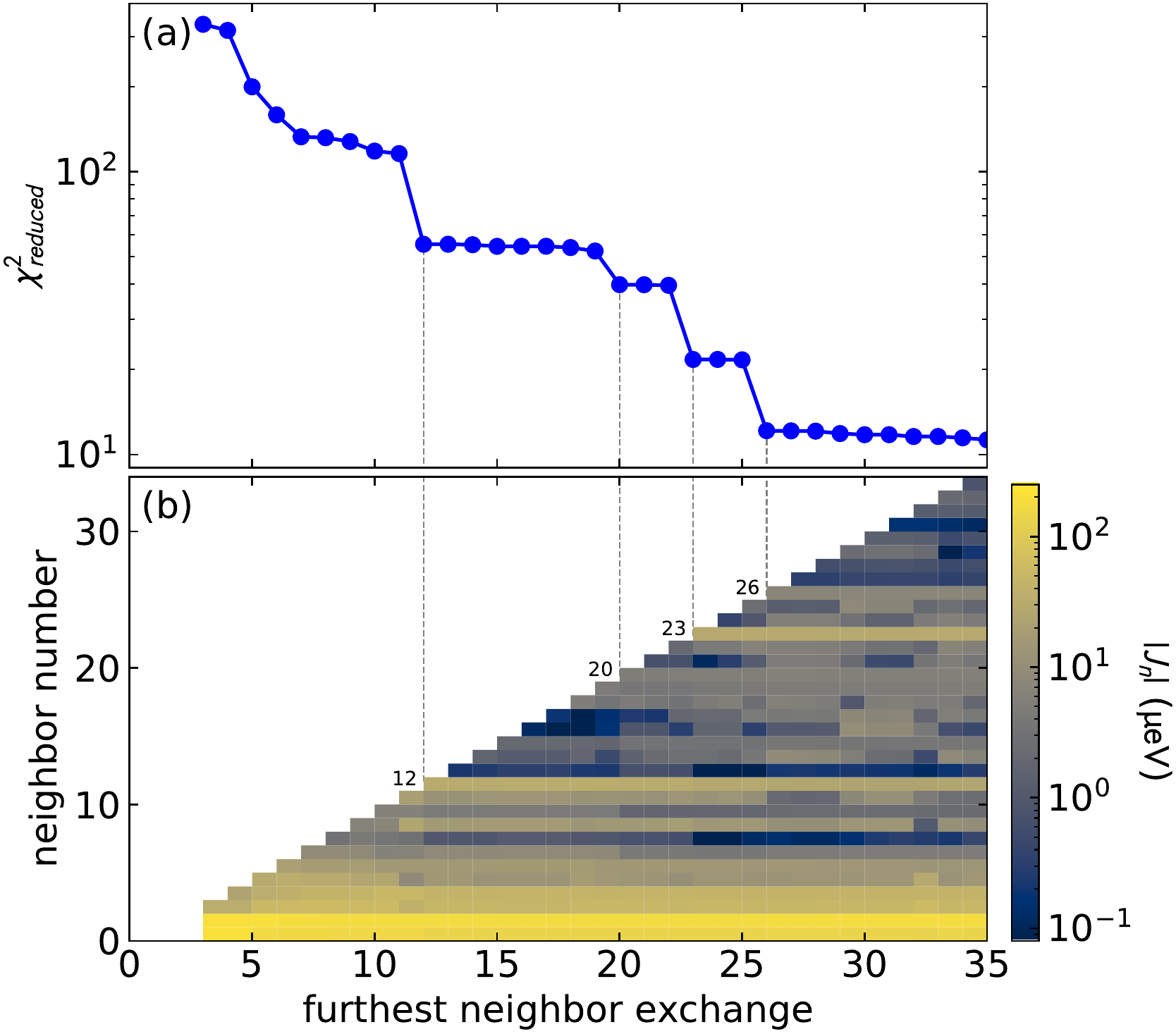}
	\caption{Fit vs number of fitted parameters. Panel (a) shows the $\chi^2_{red}$ as a function of number of parameters. Panel (b) shows the absolute value of the fitted exchange constants. Note that a sudden drop in $\chi^2_{red}$ (seen at $n = 12,\>20,\>23$, and $26$) corresponds to a well-constrained parameter. Because $\chi^2_{red}$ only slowly decreases beyond $n=26$, we cut off the fit at $n=26$.}
	\label{flo:chisqvsjn}
\end{figure}

We examined $\chi^2$ as a function of neighbors included in the model, as shown in Fig. \ref{flo:chisqvsjn}. We find that exchanges up to the 26th neighbor have a significant effect on the fit. By comparing Fig. \ref{flo:chisqvsjn} with Fig. \ref{flo:Gd_exchanges_color}, one can see that the drops in $\chi^2$ correspond to either crossing additional lattice planes as terms are added to the Hamiltonian (neighbor $n = 12,23$) or when exchanges in a particular direction not represented before are included in the Hamiltonian ($n=20,26$).
Including exchanges up to 35th neighbor continues to improve the fit, though only slightly [Fig. \ref{flo:chisqvsjn}(a)]. In our model, we ignored exchanges beyond the 26th nearest neighbor as insignificant within the uncertainty of our data. 

\begin{figure}
	\centering
	\includegraphics[width=0.98\linewidth]{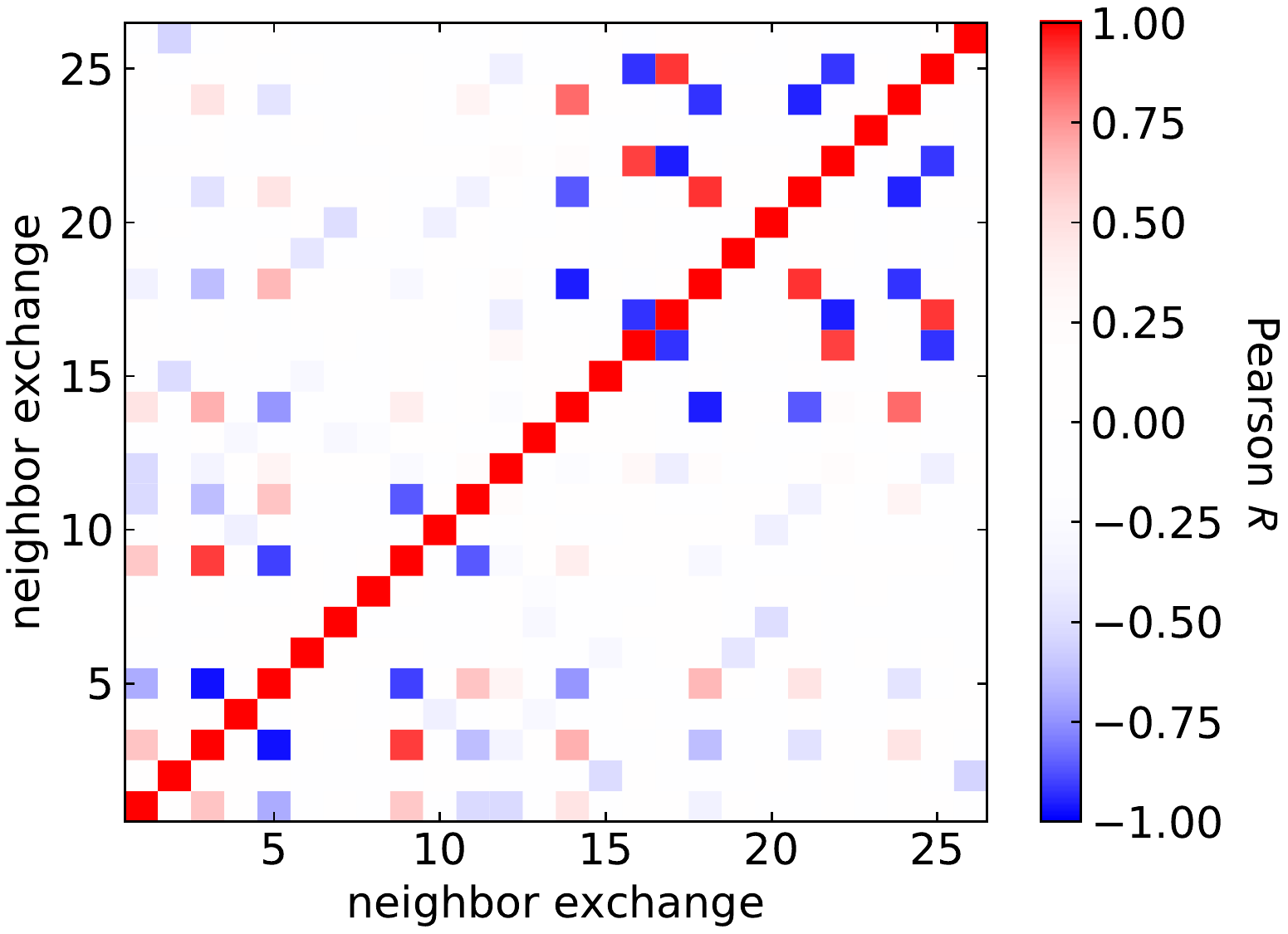}
	\caption{Correlation matrix for $J_n$ from the Gd spin wave mode fits. Red indicates positive correlation, blue indicates negative correlation. Some parameters, like $J_3$, are highly correlated with many other parameters. Others, like $J_4$, have almost no correlation with other parameters, and thus are uniquely constrained by the data.}
	\label{flo:correlationmatrix}
\end{figure}

We calculated the statistical uncertainty for these fitted parameters by computing the $\Delta \chi^2 = 1$ contour around the global minimum using a systematic search and Monte Carlo Markov chain method. This gave a reliable estimate of statistical uncertainty, as well as the correlation matrix between each parameter (visualized in Fig. \ref{flo:correlationmatrix}). As is perhaps not surprising, many of the fitted exchange constants are highly correlated with one another. Indeed, principal component analysis reveals that 78.0\% of the statistical variance of all exchange constants can be described by a single vector in $\{J_{1}, J_{2},..., J_{26}\}$. Thus, the uncertainties in Table \ref{tab:FittedJ} are by no means independent: the parameters with the largest error bars are highly correlated with a set of other exchanges.
In addition to this statistical uncertainty in the refined values, we also calculated the systematic uncertainty associated with truncating the Hamiltonian at the 26th nearest neighbor by taking the standard deviation of the parameters between $n_{max} = 27$ and $n_{max} = 35$ as the systematic uncertainty, which we added in quadrature with the statistical uncertainty to get the values in Table \ref{tab:FittedJ}. 

We compare our refined Hamiltonian to the one proposed by Lindg\aa rd \cite{Lindgard_1978} in Fig \ref{flo:SpinWaves}(i)-(l). Lindg\aa rd's Hamiltonian has a reduced $\chi^2$, $\chi^2_{red} = 442.1$ when compared to our data, whereas the Hamiltonian in Table \ref{tab:FittedJ} has $\chi^2_{red} = 12.1$. This dramatic improvement is made possible by our much larger data set.

It is worth noting that the measurements in Ref. \cite{Koehler_1970} were carried out at 78~K (compared to 5~K in our experiment). Subsequent measurements in the 1980's showed the Gd dispersion along $c$ varies greatly between 9~K and 290~K \cite{Cable_1981,Cable_1985,Cable_1989}, so some differences between the Lindg\aa rd Hamiltonian and our are expected. In general it appears that the higher frequency $Q$ modulations in the modes vanish at higher temperatures, indicating that the exchange becomes more and more short-ranged as the temperature increases \cite{Cable_1985}.

\subsection*{Comparison to RKKY}

\begin{figure}
	\centering\includegraphics[width=0.45\textwidth]{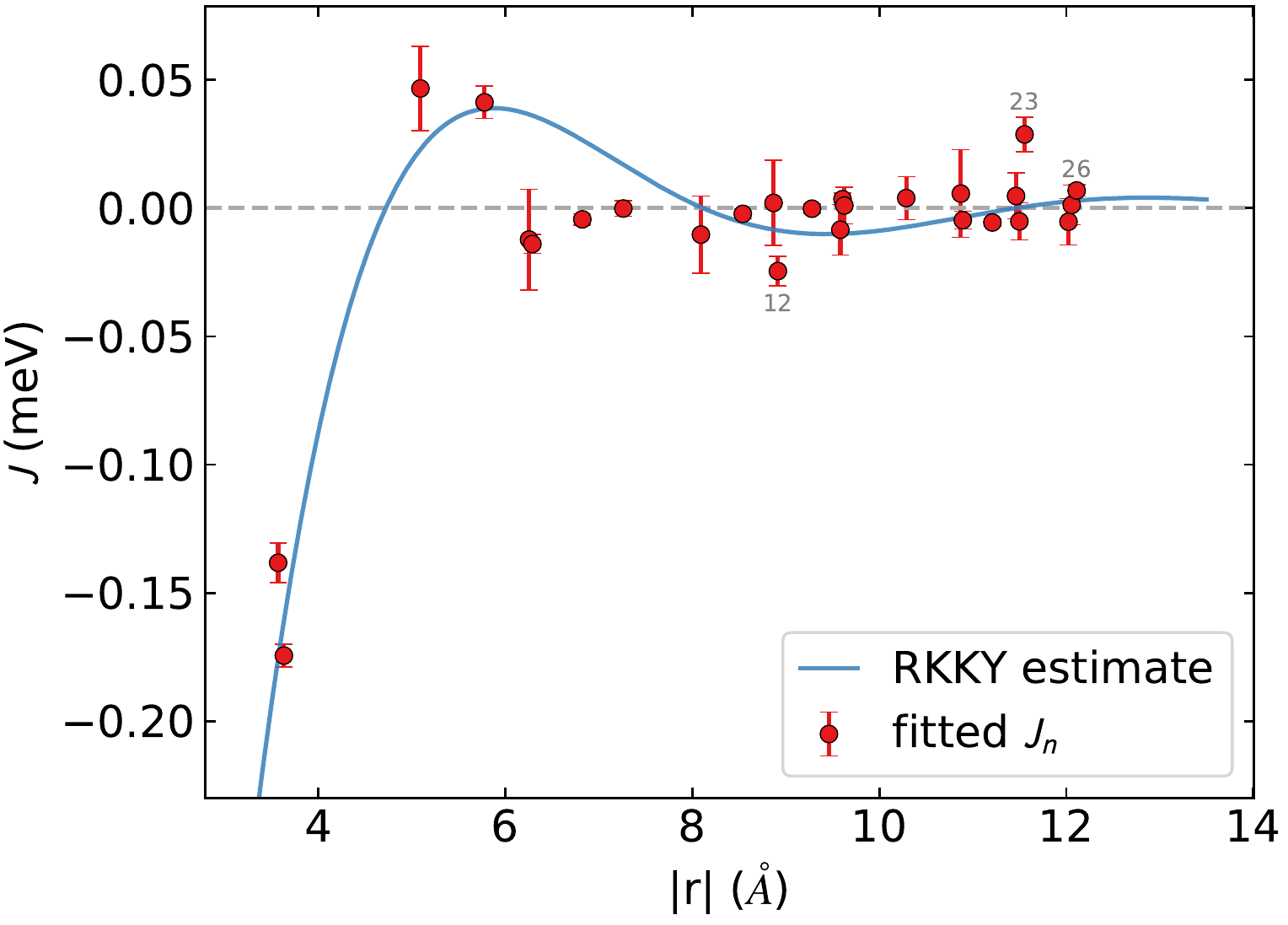}
	\caption{Refined values of magnetic exchange constants as a function of bond distance $|r|$ compared with a fitted RKKY exchange [Eq. \eqref{eq:RKKY}], see text. A few key points are labeled by their $J_n$ index.}
	\label{flo:RKKYcomparison}
\end{figure}

Despite the fact that Gd is ferromagnetic at low temperatures, many of the exchange constants are antiferromagnetic. Plotting the exchange constants against bond distance (Fig. \ref{flo:RKKYcomparison}) reveals an oscillation between ferromagnetism and antiferromagnetism as bond distance grows.
This is consistent with Ruderman-Kittel-Kasuya-Yoshida (RKKY) exchange \cite{RK_1954, Kasuya_1956, Yosida_1957}, where the magnetic exchange is mediated by conduction electrons. 

The RKKY mechanism in three dimensions predicts a magnetic exchange Hamiltonian of the form
\begin{equation}
    \mathcal{H} = A \> {\mathbf{S}_i} \cdot {\mathbf{S}_j} \big[ 2 k_\mathrm{f} r_{ij} \cos (2 k_\mathrm{f} r_{ij})  - \sin(2 k_\mathrm{f} r_{ij}) \big] / r_{ij}^4 \label{eq:RKKY}
\end{equation}
\cite{RK_1954}, 
where $k_\mathrm{f}$ is the Fermi wavevector, $r_{ij}$ is the bond distance, and $A$ is a constant. The above equation assumes a spherical Fermi surface; because this is is not quite true for Gd, we treat $k_\mathrm{f}$ in Eq. \eqref{eq:RKKY} as a fitted constant, fitting to the nearest six neighbors $J_n$ where the oscillation is clearest. This yields $k_\mathrm{f} = 0.49(3)$~\AA$^{-1}$  and the curve shown in Fig. \ref{flo:RKKYcomparison}. This fitted value is very close to the largest Fermi wavevectors measured by De Haas-Van Alphen oscillations (0.15~\AA$^{-1}$ to 0.51~\AA$^{-1}$) \cite{Mattocks_1977}, indicating that the fitted $k_\mathrm{f}$ is reasonable and close to the actual Fermi surface radius. Of course, some of the fitted $J_n$ do not follow the spherical RKKY model (most notably $J_{23}$), indicating the presence of additional exchange mechanisms (see Appendix \ref{app:RKKY} for details, where we show how purely RKKY models fail to reproduce the mode energies). Nevertheless, our results confirm that the RKKY mechanism is dominant in Gd, giving an oscillation betwteen ferromagnetic and antiferromagnetic exchange. RKKY exchange has long been thought to be dominant in Gd \cite{Turek_2003,Hindmarch_2003,Watson_1969,Lindgard_1975}, but here we show clear evidence in the exchange constants themselves.

\section{Topology and degeneracies} \label{sec:topology}

\begin{figure}
	\centering\includegraphics[width=0.45\textwidth]{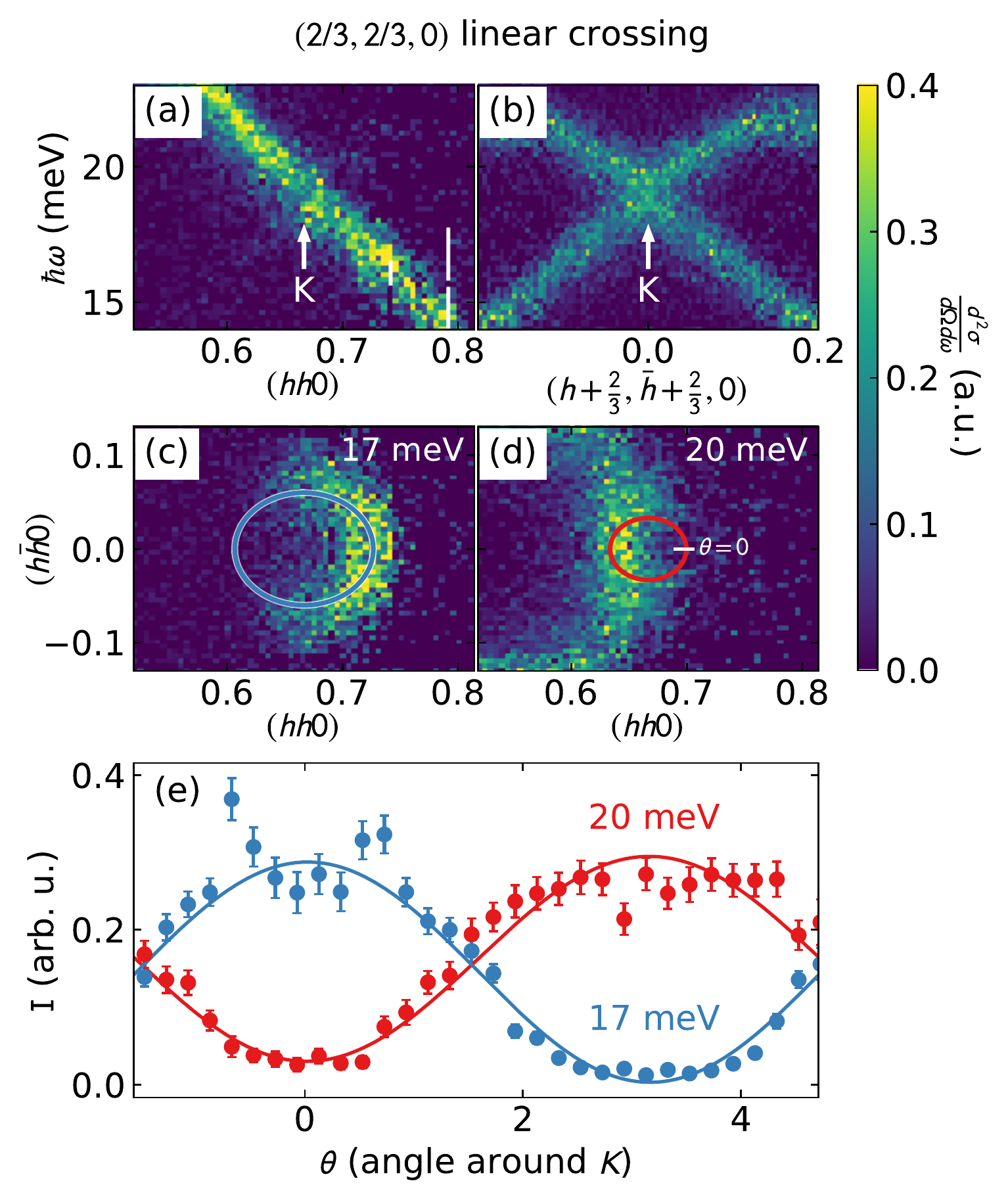}
	\caption{Linear band crossing at $K = (\frac{2}{3}\frac{2}{3}0)$. Panels (a) and (b) show the measured neutron scattering for two orthogonal cuts through $K$, highlighting the anisotropic intensity around the dispersion cone. Panels (c) and (d) show constant energy slices above and below the band crossing, showing ``intensity arcs''. Panel (e) shows the intensity binned around the circles in (c) and (d), fitted to a sin function.}
	\label{flo:Kcrossing}
\end{figure}

Some features of the Gd spin wave spectrum do not depend on the fine details of the Hamiltonian: band degeneracies \cite{Brinkman_1967,Cracknell_1970} and associated topological invariants \cite{McClarty_2021} are immune to perturbations of the exchange constants. 
As discussed in Ref. \cite{Scheie_Gd_PRL}, the Gd neutron spectrum shows  a nodal line degeneracy at $K=(1/3,1/3,\ell)$ which has a $\pi$ Berry phase around it, and a nodal plane degeneracy at half-integer $\ell$. An example of a $K$ linear band crossing is shown in Fig. \ref{flo:Kcrossing}, which has the anisotropic winding intensity characteristic of nontrivial topology \cite{McClarty_2019,shivam2017neutron}. 
The origin of the anisotropic intensity is discussed in Appendix~\ref{app:Degeneracy}.
These degeneracies are generic to all Heisenberg-only exchange Hamiltonians on the HCP lattice \cite{Brinkman_1967,Cracknell_1970}, so long as the ferromagnetic ground state is preserved.

Degeneracies can be broken by anisotropic exchange terms in the magnetic Hamiltonian. 
Magnetic dipolar exchange is one potential source of anisotropy; however in the HCP Gd lattice the lattice-summed dipolar exchange is $\approx 0.15$~meV \cite{Fujiki_1987}, which is less than 1.6\% of the lattice-summed nearest neighbor exchange strength $J_{ij} n (\frac{7}{2})^2 = 9.26$~meV, so dipolar exchange would not noticeably influence the band degeneracies in Gd.

\begin{figure}
	\centering\includegraphics[width=0.47\textwidth]{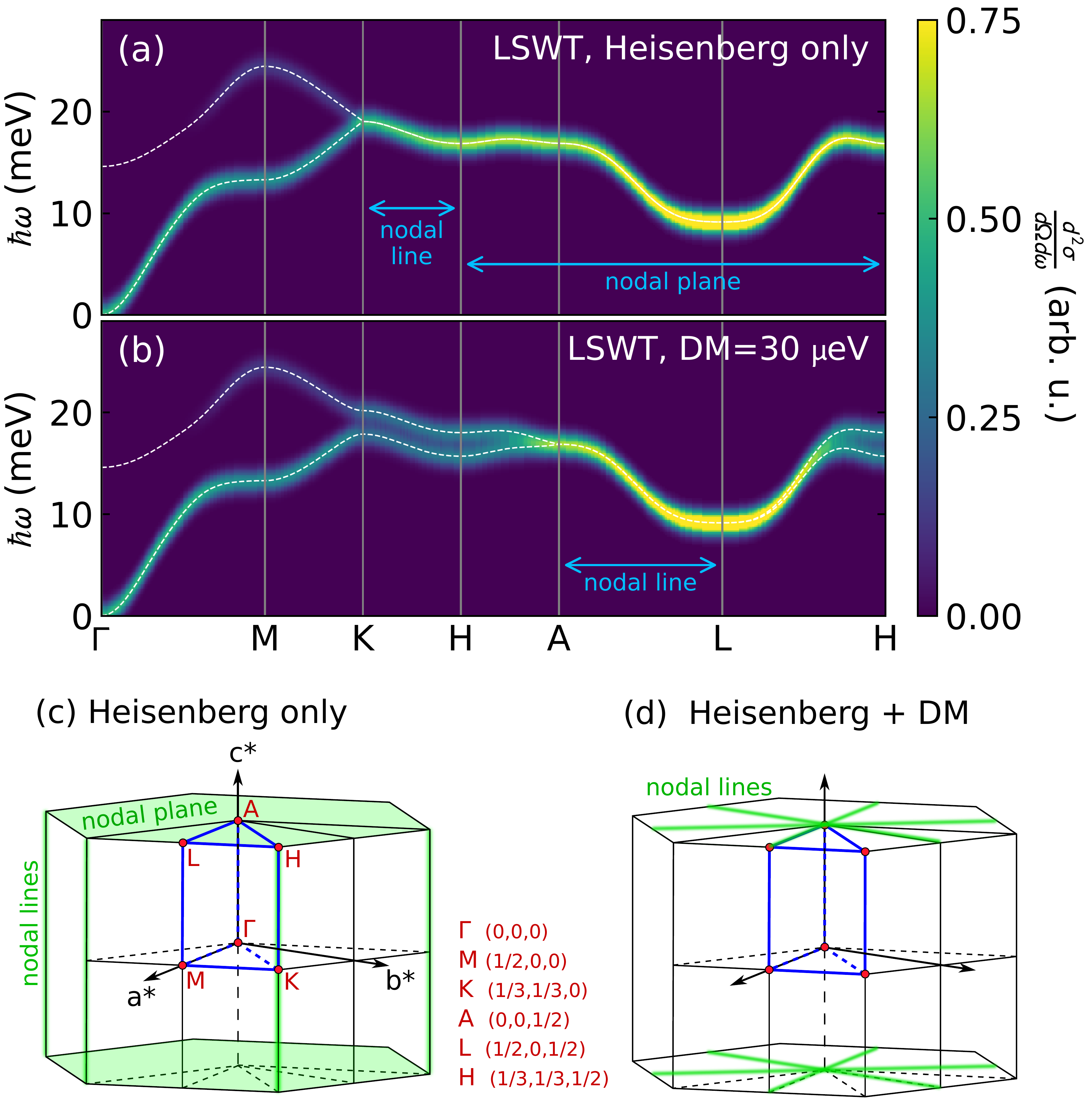}
	\caption{LSWT calculated spectrum along high symmetry directions without (a) and with (b) nonzero DM exchange interaction term on the second neighbor site. When ${\rm DM} =0$, $K$ has a linear band crossing and all $\ell = 1/2$ is a nodal plane. When ${\rm DM}> 0$, the $K \rightarrow H$ nodal line and the nodal plane degeneracies are broken, leaving nodal lines only along $A \rightarrow L$. Panel (c) shows the Brillouin zone nodal lines in (a), and panel (d) shows the same for (b).}
	\label{flo:LSWT_NodalLinePlane}
\end{figure}

In general, however, other elemental HCP magnets have unquenched orbital angular momentum, leading to off-diagonal magnetic exchange which breaks the degeneracies.
One such example is the anisotropic Dzyaloshinskii-Moriya (DM) exchange interaction
\begin{equation}
    H = \sum_{ij} \bf{D} \cdot ( \bf{S}_i \times \bf{S}_j).
\end{equation} 
Among the 30 nearest HCP neighbors, DM exchange is symmetry-allowed on the 2nd, 7th, 8th, 10th, 13th, 15th, 19th, 20th, 26th, and 27th neighbors \cite{Moriya_1960}. Each DM exchange has a similar effect on the magnon degeneracies: nonzero DM exchange will tend to lift the degeneracy at the $K$ point $(1/3, 1/3, 0)$, as well as for most of the nodal plane, as shown in Fig. \ref{flo:LSWT_NodalLinePlane}---although the splitting depends upon the spin polarization direction. 
No gap at the $\Gamma$ point $(0,0,0)$ is produced by DM exchange on the HCP lattice. However, a nodal line degeneracy along $A \rightarrow L$ is preserved even with nonzero asymmetric exchange, as shown in Fig. \ref{flo:LSWT_NodalLinePlane}. This means that asymmetric exchange leaves a triangular grid of $\ell=1/2$ nodal lines in place of the nodal plane. This grid of nodal lines is protected by residual magnetic glide symmetry $m_{1\bar{1}0} \mathcal{T}$ that exists because the DM has a U(1) symmetry.
In Gd, no mode splitting is resolvable, giving an upper bound of 3(1)~$\mu$eV on the total asymmetric DM exchange in Gd (see Appendix \ref{app:NodalPlaneDegeneracy}).

\begin{figure}
	\centering\includegraphics[width=0.48\textwidth]{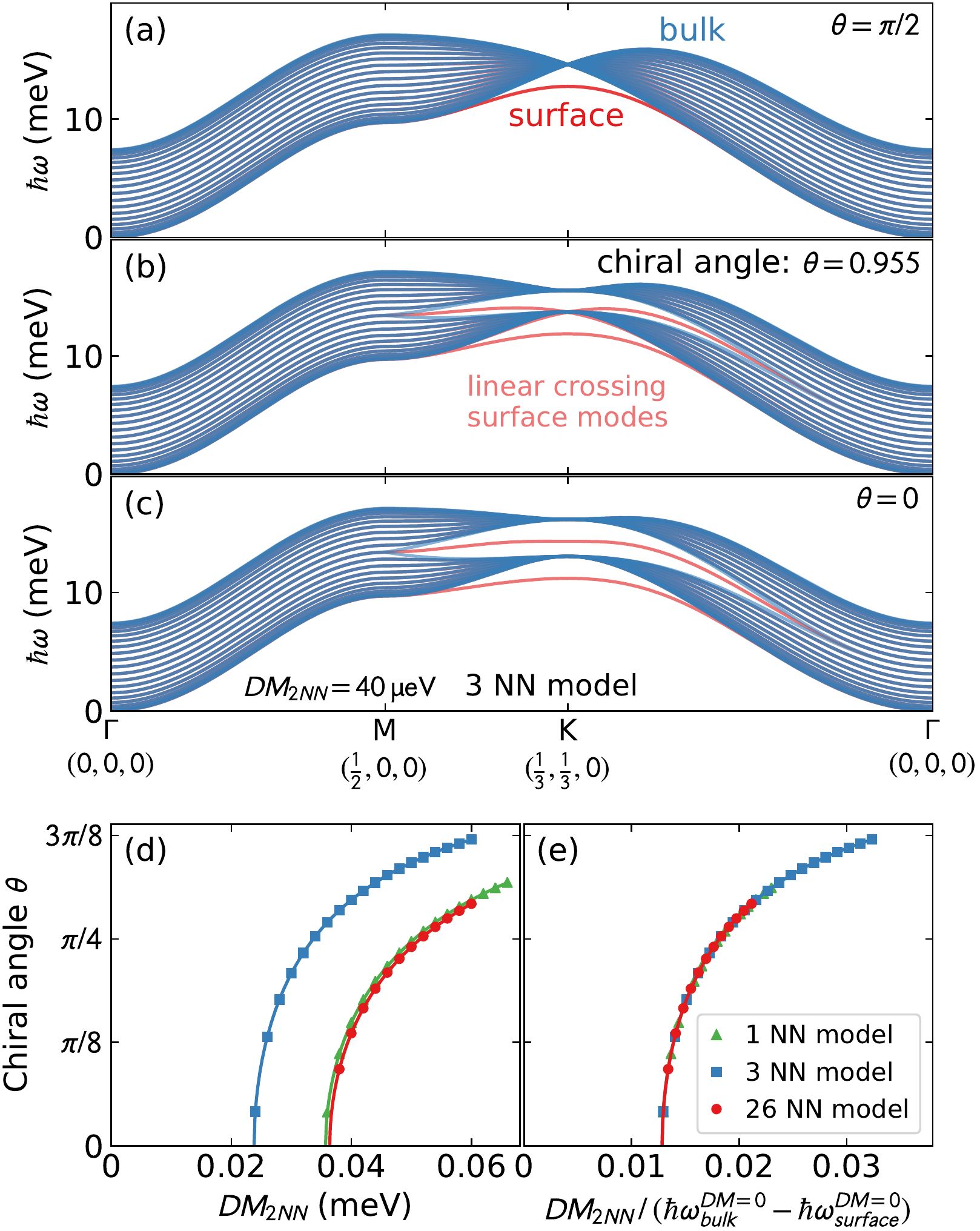}
	\caption{Chiral surface mode in the HCP ferromagnet. The top three panels show the bulk (periodic boundary conditions) and surface ($c$-axis termination) modes in a simplified 3 nearest neighbor (3NN) exchange model with a 2nd neighbor DM exchange of 40 $\mu$eV. The three panels show the spins polarized at different angles $\theta$ from the $c$ axis. The different spin polarizations induce different mode splitting because of the DM asymmetry. At a certain polarization angle, the two surface modes meet and cross with opposite slope, evidencing chiral surface magnon modes. For the simplified 3NN model, the chiral angle is $\theta = 0.955$. Panel (d) shows this chiral angle as a function of DM strength for different models: nearest neighbor exchange only (green), 3NN model (blue), and the full 26 neighbor model (red). In each case, a minimum DM exchange strength is necessary to produce the chiral mode crossing. Panel (e) shows these chiral angle is governed by the energy gap between the surface and bulk modes at $K$.}
	\label{flo:ChiralAngle}
\end{figure}

Intriguingly, the mode splitting at a given DM exchange strength can be tuned by changing the spin orientation. Because the 2nd neighbor DM vector is constrained to point along the $c$-axis, when spins are polarized in the $ab$ plane the mode splitting vanishes. Meanwhile, the mode splitting is maximal when spins are polarized along $c$. (Fig. \ref{flo:LSWT_NodalLinePlane} was calculated using the low-temperature Gd spin orientation; mode splitting with other spin orientations is explored in Fig. \ref{flo:ChiralAngle}.) Thus, in the anisotropic HCP rare earths, it may in principle be possible to tune the topological features by polarizing the spins along certain directions. 

The physical consequences of the nodal plane and the $\ell = 1/2$ nodal line grid have yet to be fully explored, but we can draw conclusions about the best understood feature: the nodal line Dirac cone at $K$.
One of the reasons that Dirac magnons are of interest is because when symmetry is broken by off-diagonal exchange, chiral surface magnons can be induced \cite{McClarty_2021}. To explore this in the HCP context, we calculated the magnons for a 20 unit cell layer supercell HCP ferromagnet using SpinW \cite{SpinW}, shown in Fig. \ref{flo:ChiralAngle}. Using a simplified three-neighbor exchange model, we calculated the modes both with periodic boundary conditions (shown in blue) and a $c$-axis termination simulated with a blank space at the top of the supercell (shown in red) with a second-neighbor DM exchange of 40 $\mu$eV. The surface modes are clearly visible as lying outside the continuum of magnon states between $\ell=0$ and $\ell=1$.

As noted above, the DM magnon mode splitting can be tuned by the spin polarization angle from the $c$-axis. The calculations in Fig. \ref{flo:ChiralAngle}(c)-(e) show that at a particular spin polarization angle, the surface magnon modes appear to cross at $K$ with opposite velocities---potentially indicating chiral edge modes \cite{McClarty_2021}. The specific angle depends upon the DM exchange interaction strength---but if DM exchange is strong enough, the system could host chiral edge states at that special polarization angle.

 We calculated this ``chiral angle'' for several different HCP models with an second-neighbor DM term in Fig. \ref{flo:ChiralAngle}(d).
For all models, the chiral angle depends on DM interaction strength as $\arccos(C/{\rm DM})$ where $C$ is a constant dependent upon the energy gap between the ${\rm DM}=0$ surface and bulk magnons at $K$. If we scale the chiral angle curve by the surface-bulk gap, the curves of various models collapse onto the same curve [\ref{flo:ChiralAngle}(e)], giving a minimum threshold value for DM exchange based on the rest of the exchange Hamiltonian. 
The situation grows more complicated for other off-diagonal exchange on the HCP lattice, but spin reorientation still generically shifts around the bands such that a special polarization direction may restore the $K$-point degeneracy.

Besides Gd, some other elemental HCP ferromagnets are dysprosium  \cite{Lindgard_1978,Nicklow_1971}, terbium  \cite{Lindgard_1978,Moller_1968}, and hexagonal ($\alpha$) cobalt \cite{perring1995high}. Unlike Gd, these materials have unquenched orbital angular momentum and anisotropic exchange. Nodal plane splitting was actually measured in Tb \cite{Moller_1968,Jensen+Mackintosh} to be $\sim 0.4$~meV---although the Tb spins being aligned in-plane \cite{Dietrich_1967} would tend to suppress mode splitting from DM, and other forms of anisotropy probably contribute to the mode splitting, rendering the actual DM strength uncertain. Be that as it may, Tb spin reorientation can be accomplished by a relatively modest $<5$~T magnetic field \cite{Roeland_1975}, which suggests the possibility of tunable topology. 

\section{Conclusion}

We have measured the magnetic spectrum of elemental Gd over the entire Brillouin zone. We fit its magnetic exchange Hamiltonian to the 26th nearest neighbor, finding that the exchange interactions approximately follow an RKKY oscillation, in good agreement with long-standing expectations that RKKY exchange is relevant to Gd. We have also explored the topology and degeneracies of Gd, using that to predict the nodal lines of anisotropic HCP ferromagnets. Finally, we have used linear spin wave simulations to suggest how tunable topology may be induced in HCP ferromagnets with DM exchange.

These results showcase the ability of modern inelastic neutron spectrometers to precisely determine the magnetic exchange interactions with rigorous uncertainty. They also reveal the magnetic behavior of an industry-critical material, providing quantities to test against future theoretical studies. Perhaps most importantly, these results show the class of HCP magnets to be a useful platform for topological magnetism.

\section*{Acknowledgments}

This research used resources at the Spallation Neutron Source, a DOE Office of Science User Facility operated by the Oak Ridge National Laboratory. The research by P.L. was supported by the Scientific Discovery through Advanced Computing (SciDAC) program funded by the US Department of Energy, Office of Science, Advanced Scientific Computing Research and Basic Energy Sciences, Division of Materials Sciences and Engineering. The work by SEN is supported by the Quantum Science Center (QSC), a National Quantum Information Science Research Center of the U.S. Department of Energy (DOE).

\appendix

\section{Experiment details}\label{app:ExpDetails}

For the SEQUOIA neutron measurements, we set the $T0$ chopper at 90~Hz, Fermi 1 chopper at 120~Hz, Fermi 2 chopper at 360~Hz for $E_i = 50$~meV, slits 36 mm wide and 14 mm tall. For $E_i = 100$~meV the same chopper speeds were used but with the Fermi 2 chopper at 540~Hz.

\section{Fitting procedure} \label{app:FittingProcedure}

The spin wave dispersions were fitted to the data from 37 different $Q$ vs $\hbar \omega$ slices. To avoid local minima from an overconstrained fit, we fit random subsets of the data, selecting between five and ten slices and optimizing the Hamiltonian using Scipy's implementation of Powell's method \cite{PowellsMethod}. Then, the best fit values were accepted or rejected based on improvement of the global $\chi^2$ (of all slices) with a simulated annealing method. This approach proved to be very effective at avoiding local $\chi^2$ minima, which are plentiful with this data set. As a final step, we fit $J_n$ to all slices in a steepest descent, which yielded the best fit parameters in Table \ref{tab:FittedJ}. This fitting procedure was repeated multiple times, and always converged to the same solution provided the annealing step was run for enough iterations.
A sample of code used for this fitting procedure, along with all the Gd scattering data, can be found at https://doi.ccs.ornl.gov/ui/doi/347. The full set of data points used in the fitting procedure is plotted in the Supplemental Information \cite{SuppMat}.

\begin{figure*}
	\centering
	\includegraphics[width=0.98\linewidth]{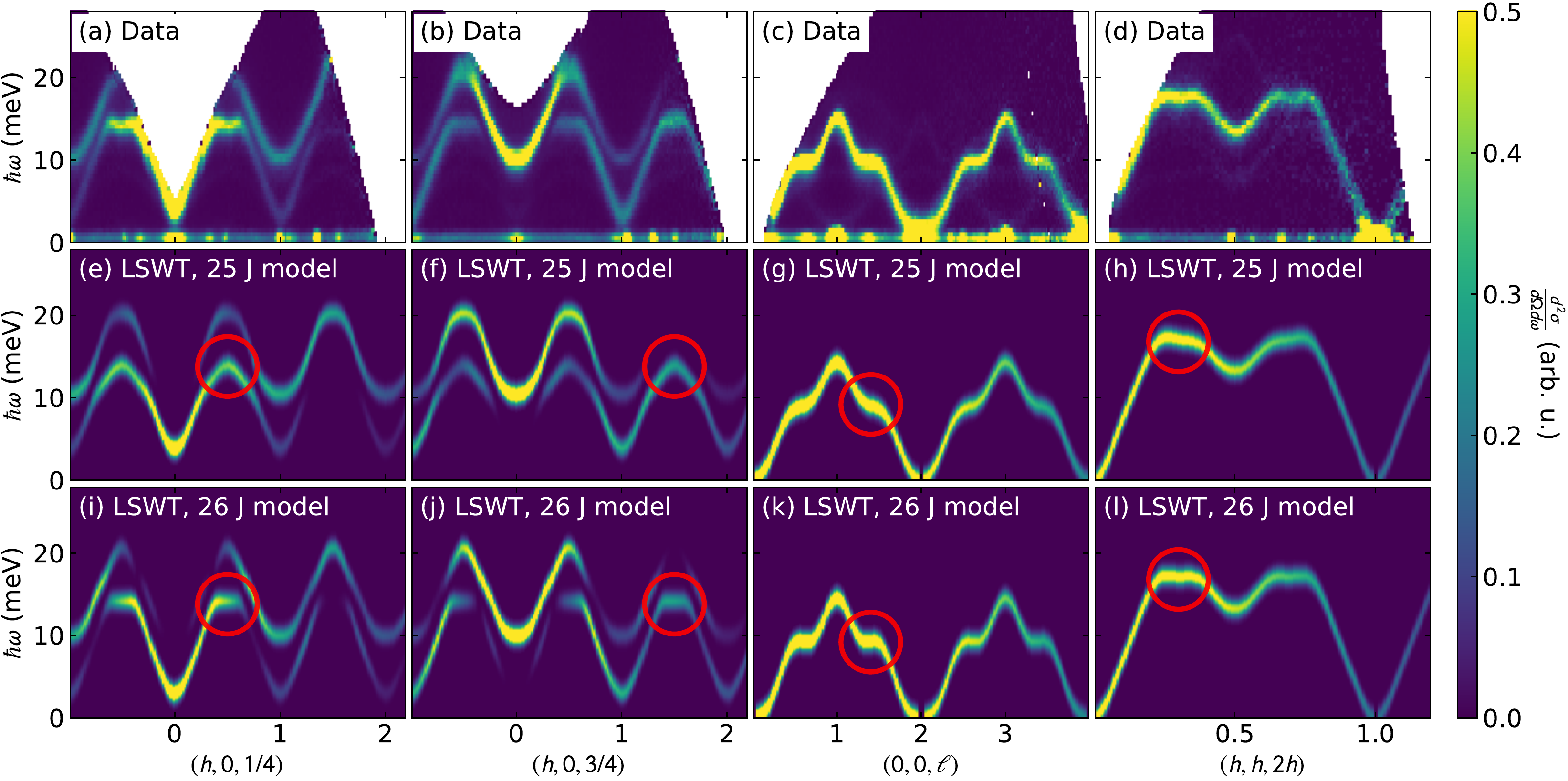}
	\caption{Comparison between experimental Gd scattering (top row) and the LSWT calculated scattering from a 25 neighbor model (middle row) and a 26 neighbor model (bottom row). As the red circles indicate, certain features in the data require the 26th neighbor $J$ to reproduce.}
	\label{flo:25to26}
\end{figure*}

As shown in Fig. \ref{flo:chisqvsjn}, the addition of certain $J_n$ in the model dramatically improves the fit. The furthest neighbor $J_n$ where we found such an effect (up to $n=35$) was $J_{26}$. To demonstrate that the 26th neighbor is indeed necessary to describe the Gd spin wave dispersion, we plot the experimental Gd scattering compared to the LSWT calculated spectra from both a 25 neighbor model and 26 neighbor model in Fig. \ref{flo:25to26}.

\begin{figure*}
	\centering
	\includegraphics[width=\linewidth]{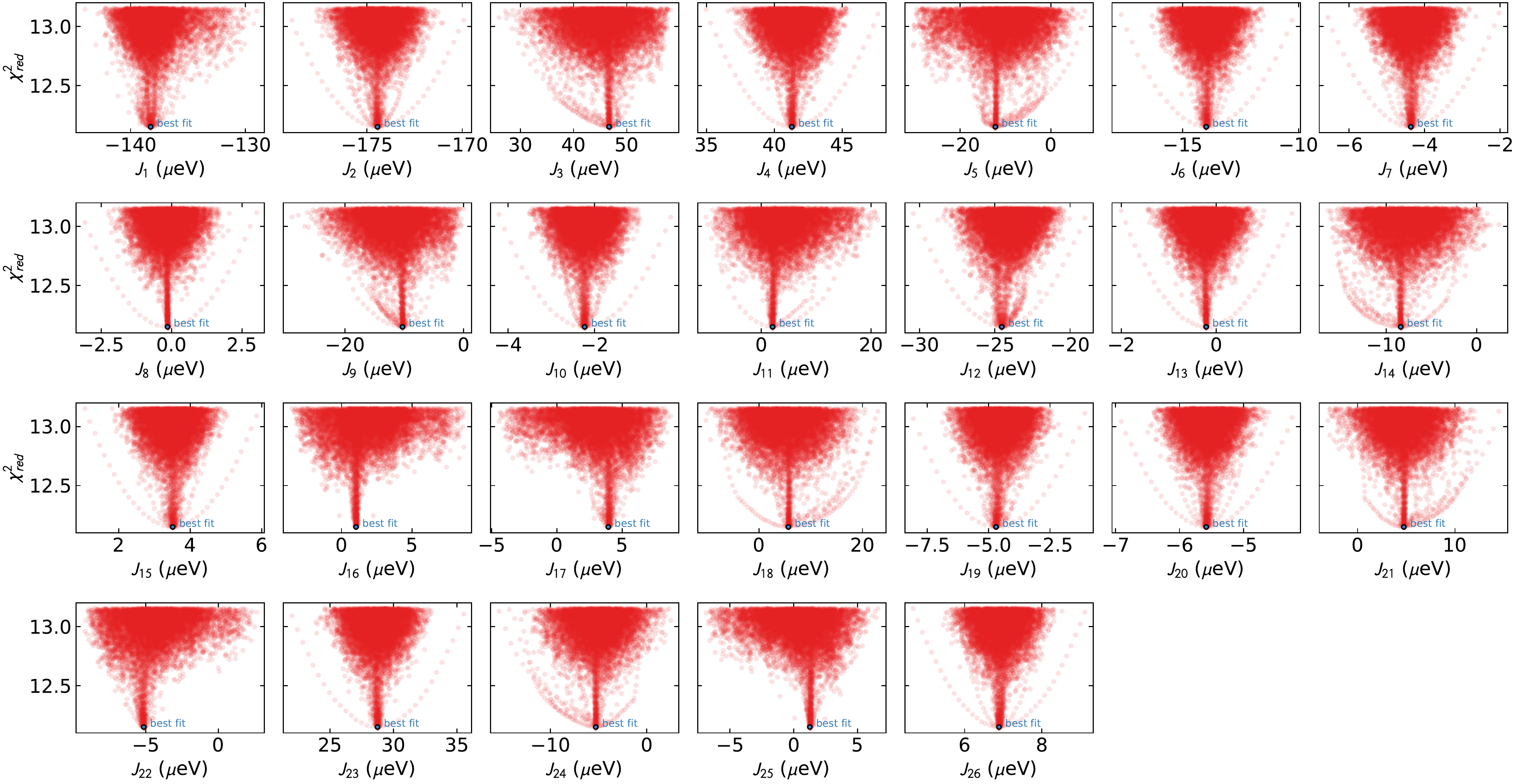}
	\caption{Possible Hamiltonian solutions within $\Delta \chi^2_{red}=1$ of the global optimum fit generated by MCMC (see text). Each panel shows the range of such solutions, which we take as an estimate of uncertainty. The small blue circle represents the best fit values.}
	\label{flo:jrangeplot}
\end{figure*}

Once the global minimum had been found, we calculated the $\chi^2$ contour by randomly sampling points around the best fit $\{J_n\}$, and keeping those whose $\chi^2_{red}$ was increased by less than one above the global optimum $\chi^2_{red}$. (This corresponds to an uncertainty of one standard deviation \cite{NumericalRecipes}.) After a few points had been identified, we used Scikit principal component analysis \cite{scikit} to identify the principal components  and accompanying standard deviations in $\{J_n\}$ along which to sample, running a Monte Carlo Markov Chain (MCMC) to collect more solutions within $\Delta \chi^2_{red} =1$. 
This approach, which yielded $\sim 6000$ possible solutions within the $\Delta \chi^2_{red} =1$ contour, allows us to sample the full $\chi^2_{red}$ landscape regardless of variation in local curvature. We take the range of possible values under the $\Delta \chi^2_{red}=1$ threshold, shown in Fig. \ref{flo:jrangeplot},  as an estimate of the statistical uncertainty of each fitted value.

\begin{figure*}
	\centering
	\includegraphics[width=\linewidth]{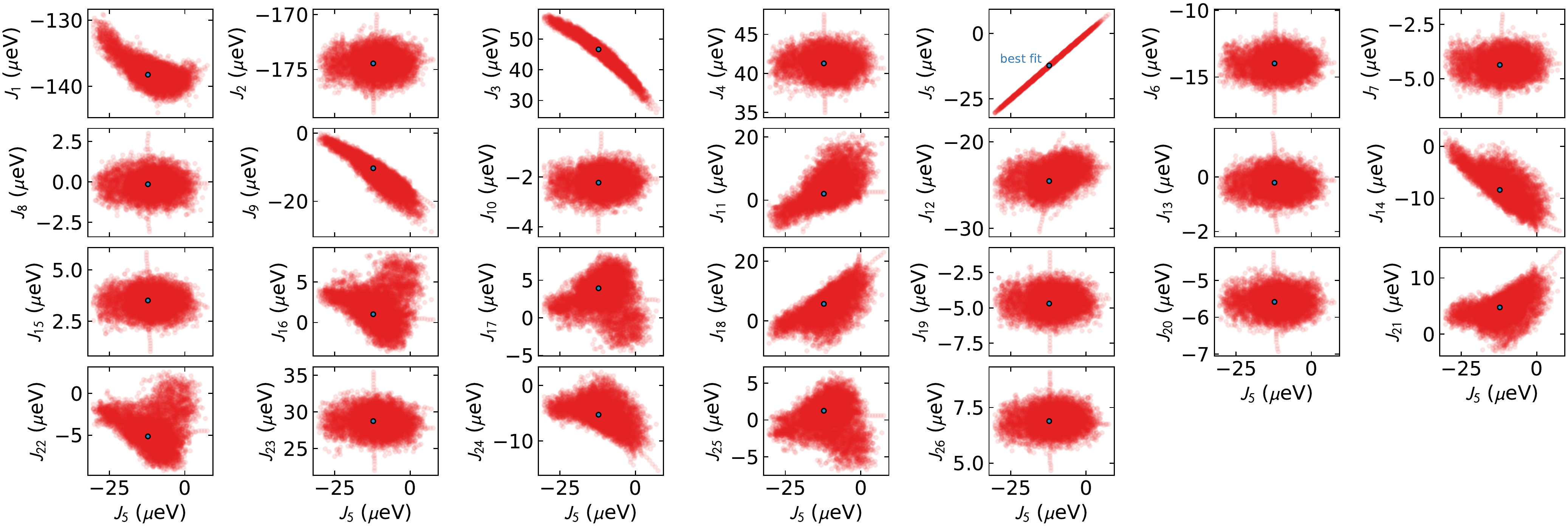}
	\caption{Hamiltonian solutions within $\Delta \chi^2_{red}=1$ plotted against $J_5$, as an example of correlations between fitted parameters. The small blue circle represents the best fit values. A circular distribution indicates no correlation, but an ellipsoidal distribution indicates high correlation. $J_1$, $J_9$, and $J_{11}$ all are highly correlated with $J_5$, and the correlations are not quite linear.}
	\label{flo:correlationjplot}
\end{figure*}

The MCMC $\chi^2_{red}$ sampling also reveals the correlation between different fitted parameters, as shown in Fig. \ref{flo:correlationjplot}. We quantify this by calculating a correlation matrix via Pearson correlation coefficients, which is visually plotted in Fig. \ref{flo:correlationmatrix}. This reveals families of highly correlated exchange constants, especially $J_3$, $J_5$, $J_9$, and $J_{11}$.

\section{RKKY calculated scattering}\label{app:RKKY}

\begin{figure*}
	\centering
	\includegraphics[width=\linewidth]{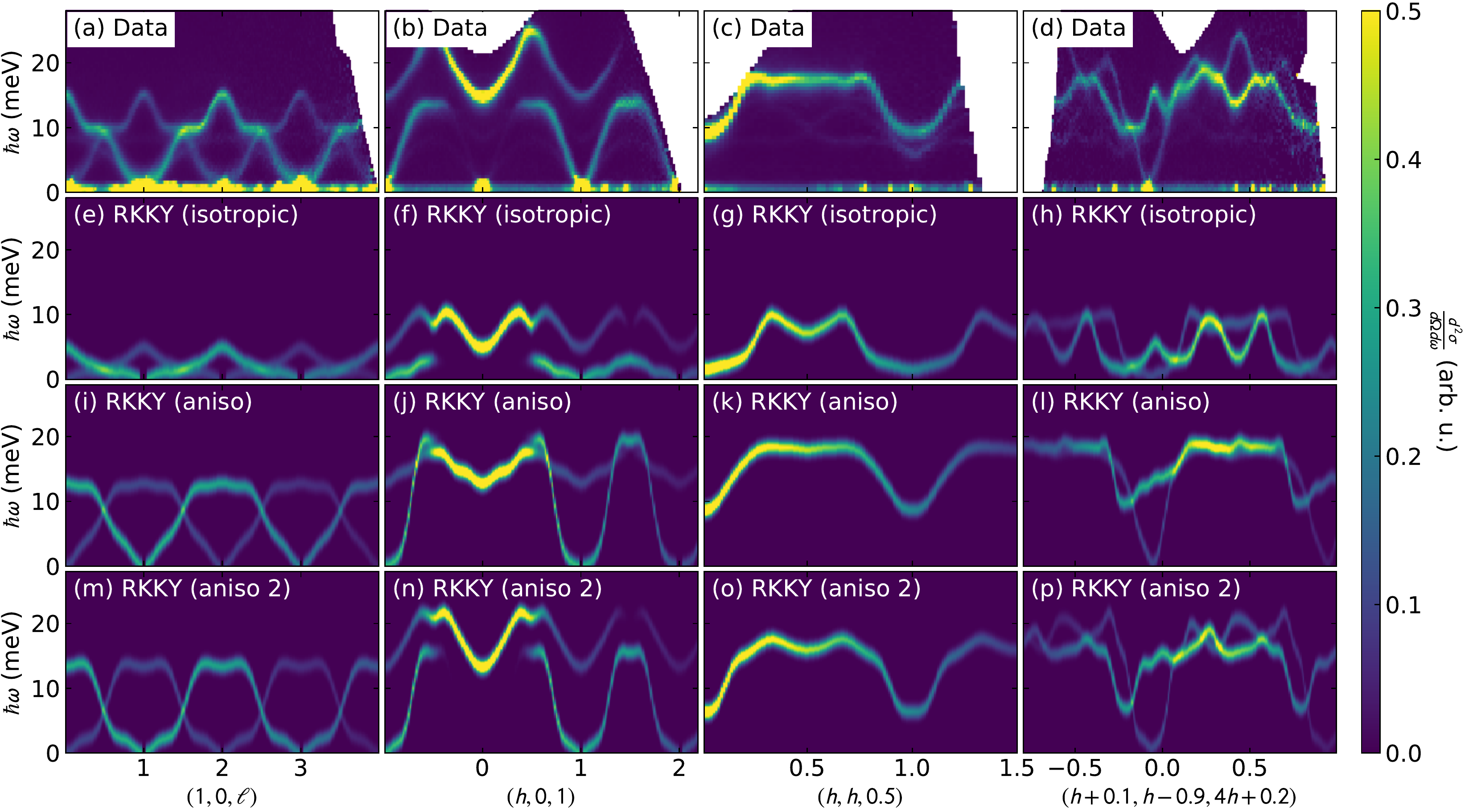}
	\caption{Comparison between experimental Gd scattering (top row) and LSWT calculated scattering from three simple RKKY models: an isotropic RKKY model (second row), an anisotropic RKKY model (third row) where the Fermi wavevector is allowed to vary as a function of bond angle from $c$, and another anisotropic RKKY model (bottom row) where the Fermi wavevector and energy scale are allowed to vary as a function of bond angle from $c$. The last model comes closest, but is still far from reproducing the details of the exchange model.}
	\label{flo:RKKY_SpinWaves}
\end{figure*}

Because the Gd magnon dispersion roughly follows an RKKY behavior, it is worth considering how close it is to an RKKY-only model. To this end, we consider three different possibilities. First, we consider an isotropic RKKY model based on Eq. \eqref{eq:RKKY}. Second, we consider an anisotropic RKKY based on Eq. \eqref{eq:RKKY} where $k_f$ varies as a function of bond angle from the $c$ axis (effectively giving three fitted parameters: an overall scale factor, $k_f^{xy}$, and $k_f^{z}$). Third, we consider an anisotropic RKKY model (``aniso 2'') based on Eq. \eqref{eq:RKKY} with different scale factors 
$$
\begin{aligned}
    \mathcal{H} = A_{xy} \> {\bf S_i} \cdot {\bf S_j} \big[ 2 k^{xy}_\mathrm{f} r_{ij} \cos (2 k_\mathrm{f} r_{ij})  - \sin(2 k^{xy}_\mathrm{f} r_{ij}) \big]r^{xy}_{ij} / r_{ij}^5  \\
    + A_z \> {\bf S_i} \cdot {\bf S_j} \big[ 2 k^{z}_\mathrm{f} r_{ij} \cos (2 k_\mathrm{f} r_{ij})  - \sin(2 k^{z}_\mathrm{f} r_{ij}) \big]r^{z}_{ij} / r_{ij}^5
\end{aligned}
$$ which gives a total of four fitted parameters: $A_{xy}$, $A_{z}$, $k_f^{xy}$, and $k_f^{z}$. These three models were each fitted to the experimental data and are plotted against the experimental data in Fig. \ref{flo:RKKY_SpinWaves}. As is evident from the plots, none of the models describe the experimental scattering very well (for the isotropic model, $\chi^2 = 1706$; for the first anisotropic model, $\chi^2 = 1467$, for the second anisotropic model, $\chi^2 = 749$). Thus a simple isotropic or ellipsoidal RKKY model is insufficient for describing the magnetic exchange of Gd.

\section{Origin of the anisotropic intensity at the band crossing}\label{app:Degeneracy}

In this section, we explain the origin of the anisotropic neutron intensity pattern at energies above and below the $K$ point linear band crossing \cite{shivam2017neutron}. The neutron structure factor for magnetic correlations is
\begin{equation}
    {\cal S}^{\alpha \beta}({\bf Q},\omega) = \sum_{\bf \ell} \int_{-\infty}^{\infty} \langle S_{0}^{\alpha}(0) S_{\bf \ell}^{\beta}(t) \rangle e^{-i {\bf Q} \cdot {\bf \ell}} e^{-i \omega t} dt,
    \label{eq:structurefactor}
\end{equation}
where $\sum_{\bf \ell}$ is a sum over all neighbor distances, $\bf Q$ is the scattering vector, and $S_{\bf r}^{\beta}(t)$ is the $\beta$ component of magnetization at site $\bf r$ at time $t$. This we may compute within linear spin wave theory. For the two band model under consideration here, the magnon wavefunctions take the form 
\begin{equation}
\psi_{\mathbf{k}\pm} = (1/\sqrt{2}) \left( \begin{array}{c} \pm \exp(i\phi_{\mathbf{k}}) \\ 1 \end{array} \right)
\end{equation} 
where $+$ is for the upper band. The momentum dependent phase enters into the dynamical structure factor as  ${\cal S}^{\alpha \beta}({\bf k},\omega) \sim 1\pm \cos \phi_{\mathbf{k}}$. For concreteness, we consider the vicinity of the Dirac crossing at the $(1/3,1/3,0)$ $K$ point, $\mathbf{k}=(1/3,1/3,0)+\delta \mathbf{k}$, the phase is the complex phase associated with $\delta k_\parallel + i\delta k_\perp$ where $\delta\mathbf{k} \equiv \delta k_\parallel (\sqrt{3}/2,1/2)+\delta k_\perp (-1/2,\sqrt{3}/2)$. This phase winds around the Dirac point. It equals zero on the far side of the Dirac point along $(h,h,0)$ from the $\Gamma$ point leading to a maximum of the intensity and $\pi$ on the near side leading to a minimum. Since the phase rotates by $\pi$ from the upper to the lower band for fixed $\mathbf{k}$, the intensity is continuous moving along $\mathbf{k}$ in the $(h,h,0)$ direction passing smoothly from the lower band to the upper band on passing through the Dirac point. 
It follows that the character of the magnon modes varies continuously in passing from the acoustic to the optical mode through the Dirac point in a given direction in momentum space.

\begin{figure*}
	\centering
	\includegraphics[width=0.98\linewidth]{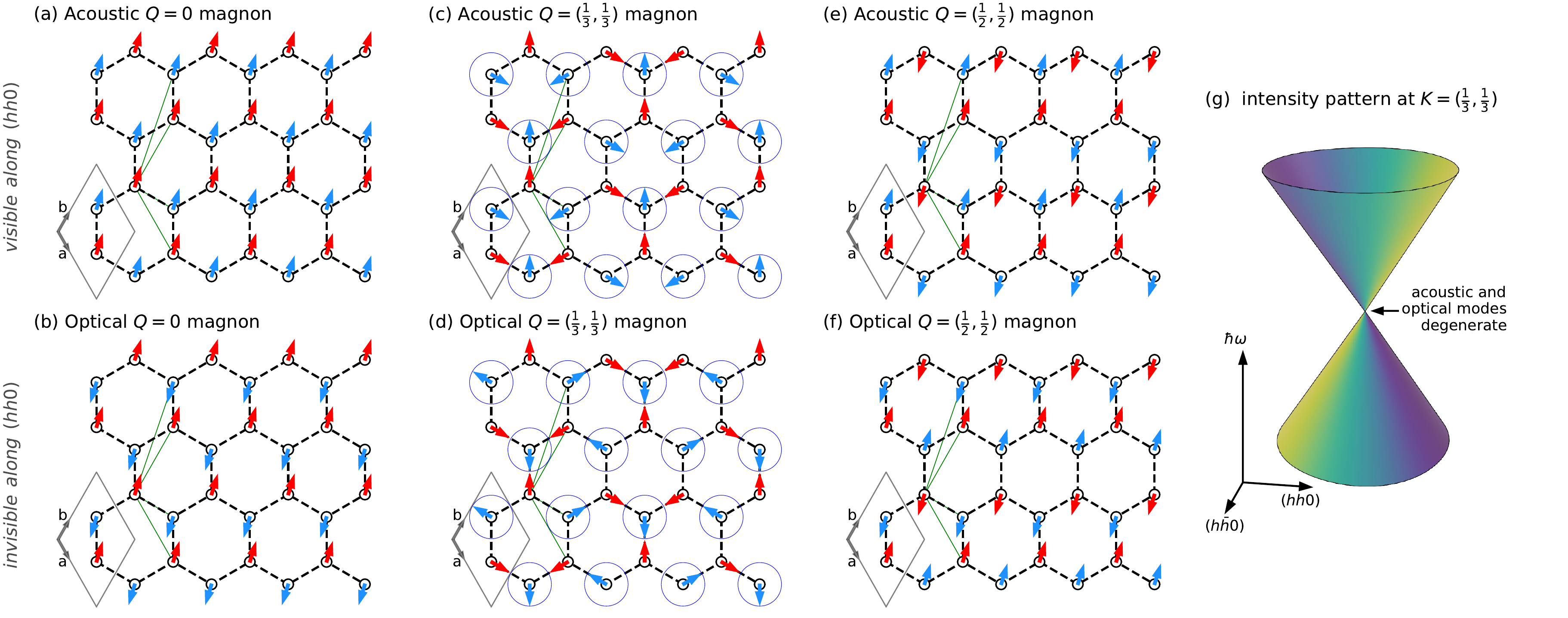}
	\caption{Origin of the anisotropic intensity around the $K$-point linear band crossing. Panels (a) and (b) show acoustic and optical modes of a $Q=0$ magnon, where the arrows indicate the spin displacement from equilibrium. Panels (c) and (d) show acoustic and optical modes of a $Q=(\frac{1}{3}, \frac{1}{3})$ magnon; (e) and (f) show the same for $Q=(\frac{1}{2}, \frac{1}{2})$ magnon. With only Heisenberg magnetic exchange, these two modes are related by a degenerate global rotation of a single sublattice at $Q=(\frac{1}{3}, \frac{1}{3})$, highlighted by the dark blue circles. Meanwhile, the acoustic mode becomes lower energy at $Q=(\frac{1}{2}, \frac{1}{2})$, signaling a mode crossing between $Q=0$ and $Q=(\frac{1}{2}, \frac{1}{2})$ which occurs at $Q=(\frac{1}{3}, \frac{1}{3})$. The green lines indicate the $\bf \ell$ in Eq. \eqref{eq:structurefactor} which give an equivalent $e^{-i {\bf Q} \cdot {\bf \ell}}$ for a neutron scattered along $(hh0)$. Because of the staggered spin displacements, the optical modes are always invisible, leading to the anisotropic intensity pattern in panel (e), where  the color coding on the double cone corresponds to $S({\bf Q}, \omega)$.}
	\label{flo:honeycombKcrossing}
\end{figure*}

This can be visualized following the ``physical picture of a spin wave'' in Squires \cite{Squires}. 
In the acoustic mode, where two spins in the unit cell cant the same direction and the optical mode the two spins cant in opposite directions (Fig. \ref{flo:honeycombKcrossing}). When $\bf Q$ is along the $(hh0)$ direction in the bipartite hexagonal lattice, something peculiar happens: the optical magnon mode has zero intensity. This is because each $\langle S_{0}^{\alpha}(0) S_{\bf \ell}^{\beta}(t) \rangle$ has a corresponding $\langle S_{0}^{\alpha}(0) S_{\bf \ell}^{\beta}(t) \rangle$ of the opposite sign with the same value of ${\bf Q} \cdot {\bf \ell}$ when $\bf Q$ is along the $(hh0)$. This is visually clear from Fig. \ref{flo:honeycombKcrossing}(b) and (d). Thus, when the sum $\sum_{\bf \ell}$ is carried out, ${\cal S}^{\alpha \beta}({\bf Q},\omega) = 0$ for the optical mode. 
Note that this is only true when ${\bf Q}=(hh0)$. As soon as ${\bf Q}$ gains any orthogonal components, the optical mode gains nonzero intensity.
Assuming net ferromagnetic Heisenberg interactions on the hexagonal lattice, at $Q=0$ the acoustic mode is zero energy and the optical mode has finite energy.
At the $K=(\frac{1}{3}, \frac{1}{3})$ point with Heisenberg exchange, the acoustic and optical magnon bands become perfectly degenerate. This is visually illustrated in Fig. \ref{flo:honeycombKcrossing}(c)-(d), where a global rotation on a single sublattice does not change the overall system energy. Moving along $(hh)$ further, at $Q=(1/2,1/2)$ the optic mode has lower energy than the acoustic mode [Fig. \ref{flo:honeycombKcrossing}(e)-(f)], as the spin canting is more along the nearest neighbor directions with the optic at $Q=(1/2,1/2)$. Thus, at $K=(\frac{1}{3}, \frac{1}{3})$ the modes linearly cross, meaning the``upper branch'' switches from zero intensity to nonzero intensity, leading to the peculiar anisotropic intensity pattern shown in panel (g).
Note that this is the case for every bipartite hexagonal system, including both the 3D HCP lattice and the 2D honeycomb lattice.

\begin{figure}
	\centering\includegraphics[width=0.45\textwidth]{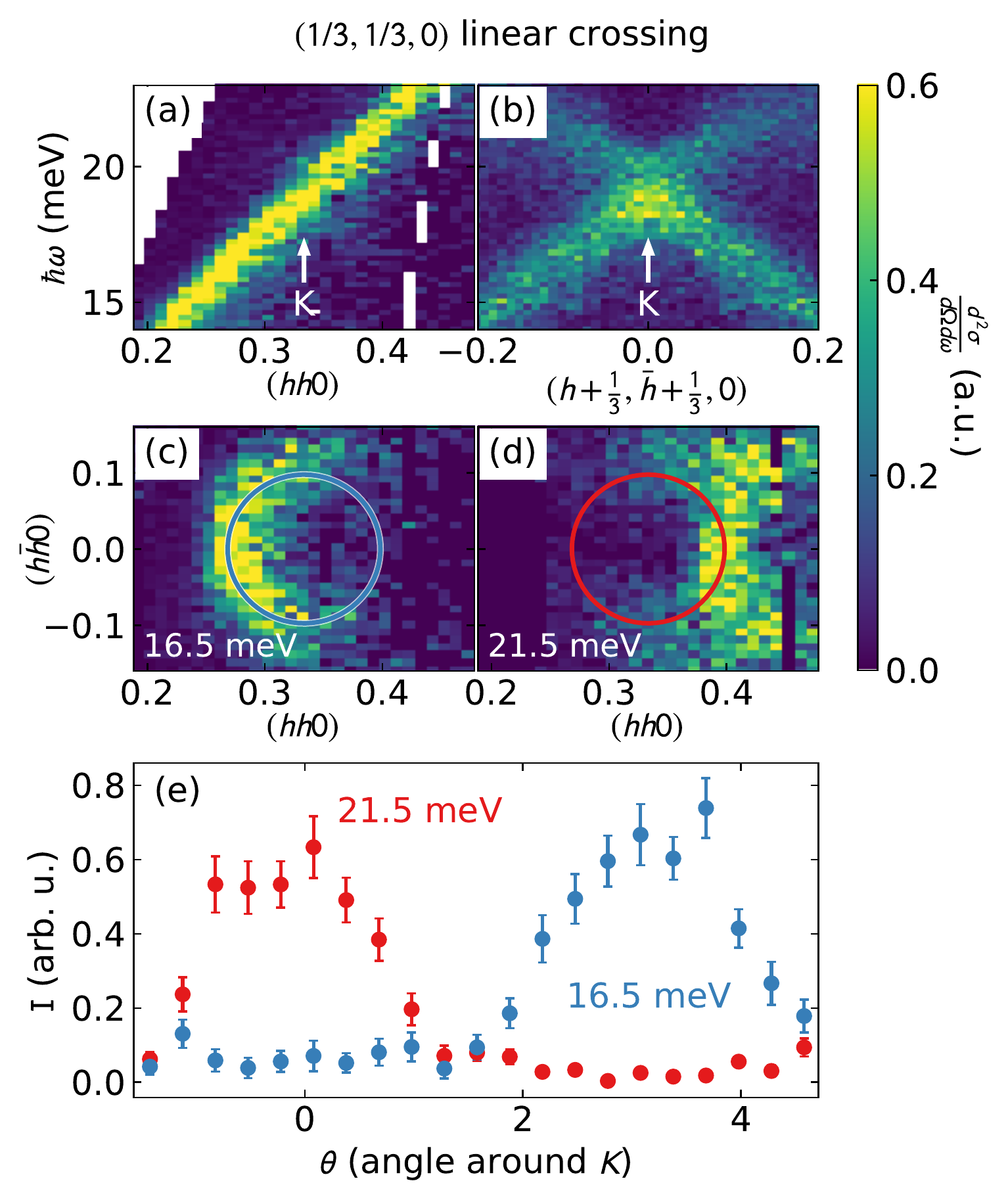}
	\caption{Linear band crossing at $K = (\frac{1}{3}\frac{1}{3}0)$ measured with $E_i = 100$~meV neutrons. The resolution and statistics are notably worse than Fig. \ref{flo:Kcrossing} with $E_i=50$~meV, but the anisotropic intensity is still visible.}
	\label{flo:Kcrossing2}
\end{figure}

This intensity pattern is observed in Gd at the $K=(2/3,2/3,0)$ point as shown in Fig. \ref{flo:Kcrossing}, though the fact that it is in the neighboring Brillouin zone means the acoustic and optic modes have swapped places, and the acoustic mode energy decreases as $hh$ increases.
Another $K$-point linear crossing in the Gd data are shown in Fig. \ref{flo:Kcrossing2}. 

\section{Nodal plane degeneracy}\label{app:NodalPlaneDegeneracy}

\begin{figure}
	\centering\includegraphics[width=0.44\textwidth]{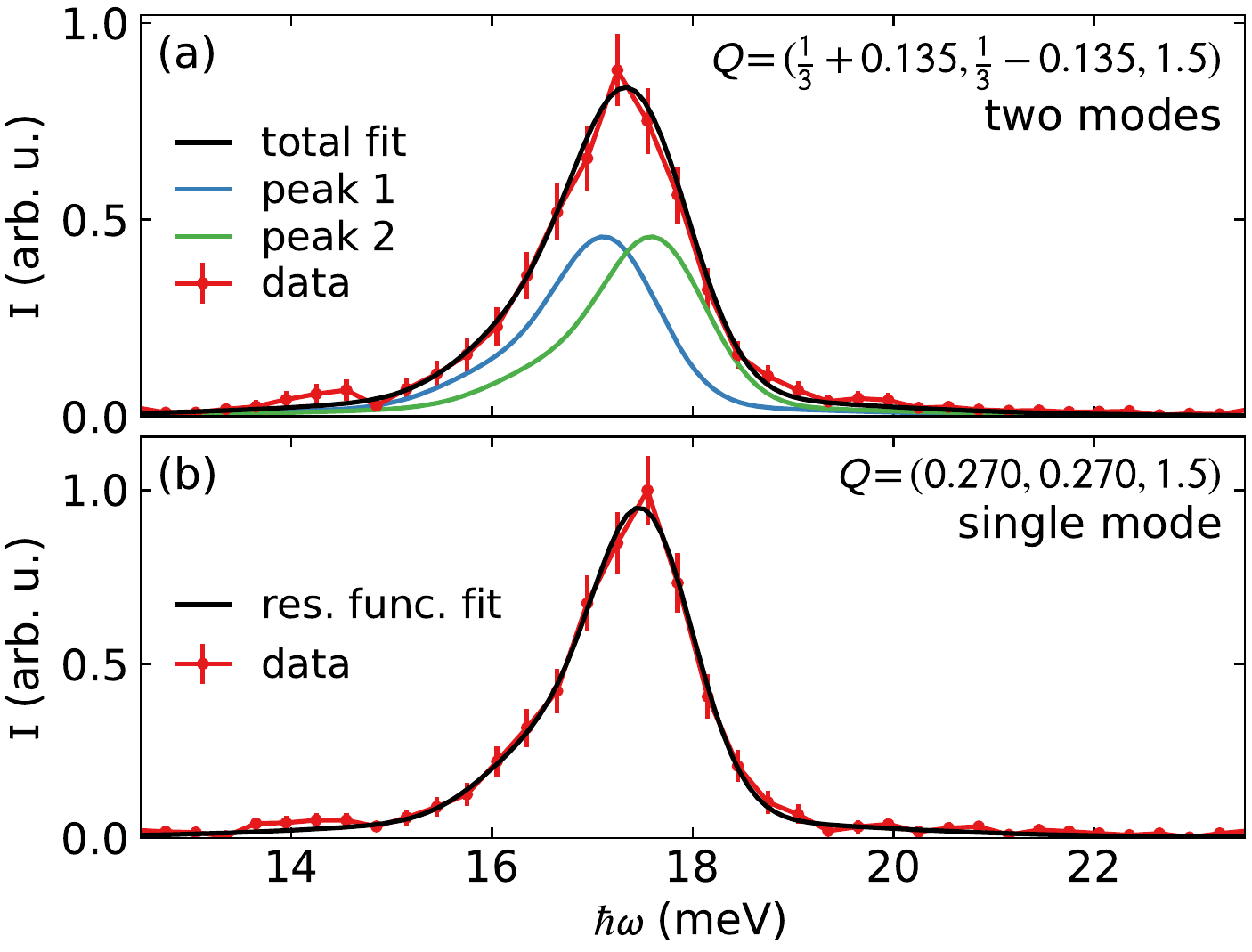}
	\caption{Mode splitting at $\ell = 3/2$ for Gd. The bottom panel shows a constant $Q$ cut on the $(h,h,3/2)$ line where only one mode has intensity. The top panel shows a constant $Q$ cut near $K$ along $(\frac{1}{3}+h,\frac{1}{3}-h,3/2)$ where both modes have intensity. Using the bottom cut to define a resolution profile, we fit two modes of equal intensity to the top cut, yielding a maximum mode splitting of 0.24(5) meV---well beyond the resolution of this experiment.}
	\label{flo:MaximalModeSplitting}
\end{figure}

To within the instrumental resolution of this experiment, the acoustic and optical magnon modes are degenerate at half-integer $\ell$ in Gd. To obtain an upper bound on the Gd mode splitting---and thus on asymmetric DM interactions---we fit the scattering at $\ell=1/2$ to a double-peak as shown in Fig. \ref{flo:MaximalModeSplitting}. Using the $(h,h,3/2)$ line where only one mode is visible to define the resolution shape and width, we find a fitted maximum double-peak splitting of 0.24(5)~meV, which corresponds to a maximum second neighbor DM exchange of 3(1)~$\mu$eV. Of course, the peak broadening observed in Fig. \ref{flo:MaximalModeSplitting}(b) could be from differences in resolution function between the two points---so this fit gives only an upper bound on the DM exchange interaction: at least 50 times weaker than the $J_2$ Heisenberg exchange interaction. (For more discussion of mode broadening see the Supplemental Information \cite{SuppMat} which includes ref. \cite{PyChop}.)

\subsection*{Symmetry and the DM interaction}

From the perspective of symmetries, we can theoretically understand the DM interaction breaking nodal plane degeneracy as follows.
The second neighbor DM with $\mathbf{D}=D\hat{\mathbf{z}}$ has spin-space symmetry. When the moments are in the plane, the $U(1)$ spin symmetry of the coupling together with physical time reversal lead to an effective time reversal $\mathcal{T}^*$ operator that guarantees the presence of the nodal plane degeneracy. If instead the moments have a nonzero out-of-plane component, the pure effective time reversal symmetry induced by spin-space symmetry is broken. For the most general anisotropic coupling the $U(1)$ will also be absent and, in this case too, effective pure time reversal symmetry is broken. A robust consequence of this symmetry breaking is the gapping out of the nodal plane and the zone corner nodal lines as shown in Fig. \ref{flo:LSWT_NodalLinePlane}.


Even though the effective time reversal symmetry is broken when the moments lie out of the xy plane, there is a residual higher symmetry when $\mathbf{D}=D\hat{\mathbf{z}}$ is present. Examining operations that leave the magnetic structure invariant we see that in-plane two-fold rotation axes and mirror planes perpendicular to the triangular planes are broken by the fact that the moments are tilted out of the xy plane. However, these symmetries survive when paired with time reversal symmetry and the $U(1)$. The is because the combination of the two-fold spin rotation in each case can be undone by rotating the moments about $z$. 

The symmetry that will be relevant to us is the set of glides times time reversal. One such magnetic glide is $m_{1\bar{1}0} \mathcal{T}$. This has the effect of swapping sublattices, taking $(k_x,k_y,k_z)\rightarrow (-k_x,k_y,-k_z)$ and complex conjugating the Hamiltonian. The action of glide on the moments is to rotate by $\pi$ about the axis perpendicular to the mirror plane; time reversal and $U(1)$ then restore the moment directions. The non-symmorphic part of the symmetry enters, as for the screw \cite{Scheie_Gd_PRL}, as an overall $\exp(i k_z)$ phase when the glide is carried out twice. Overall, on the two-band model we find
\begin{align}
{\small
\left(  \begin{array}{cc} 0 & 1 \\ e^{-ik_z} & 0  \end{array} \right)\left(  \begin{array}{cc} A(-k_x,k_y,-k_z) & B(-k_x,k_y,-k_z) \\ B^*(-k_x,-k_y,k_z) & A'(-k_x,-k_y,k_z)  \end{array} \right)\left(  \begin{array}{cc} 0 & e^{ik_z} \\ 1 & 0  \end{array} \right)}\nonumber \\  = H(k_x,k_y,k_z)
\end{align}
so $A(k_x,k_y,k_z)= A'(-k_x,k_y,-k_z)$. This implies that, on the surface $k_z=\pm\pi$ and on the invariant momentum line $k_x=0$, $A=A'$. 

In addition, the inversion symmetry and two-fold screw symmetry force $B$ to vanish on the $k_z=\pm \pi$ surface. Inversion has the effect of taking $\mathbf{k}$ to $-\mathbf{k}$ and swapping the sublattices so $B(k_x,k_y,k_z)= B^*(-k_x,-k_y,-k_z)$. The constraint from the screw symmetry on $B$ is $e^{i c k_z} B^*(-k_x,-k_y,k_z) = B(k_x,k_y,k_z)$. Taking both constraints together we find that $B$ must vanish at $k_z=\pm \pi$ as claimed.

Since $B$ vanishes and $A=A'$ on this surface, there is a degeneracy on this line. There are two further symmetries of this sort related by $C_{3z}$ operations. We have therefore shown that there are residual nodal lines on the hexagonal zone boundary when the exchange is anisotropic. 

For the most general anisotropic exchange (i.e., beyond the DM coupling considered above), time reversal symmetry survives only when the moments lie in the mirror plane. In that case the corresponding nodal line remains and the others are gapped out. For collinear moments in a nonsymmetric direction there are no degeneracies in the spectrum. The fact that nodal plane degeneracies are preserved to within experimental resolution shows that such anisotropic exchanges are negligible in Gd.

%


\begin{thebibliography}{52}%
	\makeatletter
	\providecommand \@ifxundefined [1]{%
		\@ifx{#1\undefined}
	}%
	\providecommand \@ifnum [1]{%
		\ifnum #1\expandafter \@firstoftwo
		\else \expandafter \@secondoftwo
		\fi
	}%
	\providecommand \@ifx [1]{%
		\ifx #1\expandafter \@firstoftwo
		\else \expandafter \@secondoftwo
		\fi
	}%
	\providecommand \natexlab [1]{#1}%
	\providecommand \enquote  [1]{``#1''}%
	\providecommand \bibnamefont  [1]{#1}%
	\providecommand \bibfnamefont [1]{#1}%
	\providecommand \citenamefont [1]{#1}%
	\providecommand \href@noop [0]{\@secondoftwo}%
	\providecommand \href [0]{\begingroup \@sanitize@url \@href}%
	\providecommand \@href[1]{\@@startlink{#1}\@@href}%
	\providecommand \@@href[1]{\endgroup#1\@@endlink}%
	\providecommand \@sanitize@url [0]{\catcode `\\12\catcode `\$12\catcode
		`\&12\catcode `\#12\catcode `\^12\catcode `\_12\catcode `\%12\relax}%
	\providecommand \@@startlink[1]{}%
	\providecommand \@@endlink[0]{}%
	\providecommand \url  [0]{\begingroup\@sanitize@url \@url }%
	\providecommand \@url [1]{\endgroup\@href {#1}{\urlprefix }}%
	\providecommand \urlprefix  [0]{URL }%
	\providecommand \Eprint [0]{\href }%
	\providecommand \doibase [0]{https://doi.org/}%
	\providecommand \selectlanguage [0]{\@gobble}%
	\providecommand \bibinfo  [0]{\@secondoftwo}%
	\providecommand \bibfield  [0]{\@secondoftwo}%
	\providecommand \translation [1]{[#1]}%
	\providecommand \BibitemOpen [0]{}%
	\providecommand \bibitemStop [0]{}%
	\providecommand \bibitemNoStop [0]{.\EOS\space}%
	\providecommand \EOS [0]{\spacefactor3000\relax}%
	\providecommand \BibitemShut  [1]{\csname bibitem#1\endcsname}%
	\let\auto@bib@innerbib\@empty
	\bibitem [{\citenamefont {Urbain}\ \emph {et~al.}(1935)\citenamefont {Urbain},
		\citenamefont {Weiss},\ and\ \citenamefont {Trombe}}]{Urbain_1935}%
	\BibitemOpen
	\bibfield  {author} {\bibinfo {author} {\bibfnamefont {G.}~\bibnamefont
			{Urbain}}, \bibinfo {author} {\bibfnamefont {P.}~\bibnamefont {Weiss}},\ and\
		\bibinfo {author} {\bibfnamefont {F.}~\bibnamefont {Trombe}},\ }\href@noop {}
	{\bibfield  {journal} {\bibinfo  {journal} {Comptes rendus}\ }\textbf
		{\bibinfo {volume} {200}},\ \bibinfo {pages} {2132} (\bibinfo {year}
		{1935})}\BibitemShut {NoStop}%
	\bibitem [{\citenamefont {Bauer}\ \emph {et~al.}(2011)\citenamefont {Bauer},
		\citenamefont {Diamond}, \citenamefont {Li}, \citenamefont {McKittrick},
		\citenamefont {Sandalow},\ and\ \citenamefont {Telleen}}]{chu2011critical}%
	\BibitemOpen
	\bibfield  {author} {\bibinfo {author} {\bibfnamefont {D.}~\bibnamefont
			{Bauer}}, \bibinfo {author} {\bibfnamefont {D.}~\bibnamefont {Diamond}},
		\bibinfo {author} {\bibfnamefont {J.}~\bibnamefont {Li}}, \bibinfo {author}
		{\bibfnamefont {M.}~\bibnamefont {McKittrick}}, \bibinfo {author}
		{\bibfnamefont {D.}~\bibnamefont {Sandalow}},\ and\ \bibinfo {author}
		{\bibfnamefont {P.}~\bibnamefont {Telleen}},\ }\href
	{https://www.energy.gov/sites/default/files/DOE_CMS2011_FINAL_Full.pdf}
	{\bibinfo {title} {Critical materials strategy}} (\bibinfo {year}
	{2011})\BibitemShut {NoStop}%
	\bibitem [{\citenamefont {Sch{\"u}ler}\ \emph {et~al.}(2011)\citenamefont
		{Sch{\"u}ler}, \citenamefont {Buchert}, \citenamefont {Liu}, \citenamefont
		{Dittrich},\ and\ \citenamefont {Merz}}]{schuler2011study}%
	\BibitemOpen
	\bibfield  {author} {\bibinfo {author} {\bibfnamefont {D.}~\bibnamefont
			{Sch{\"u}ler}}, \bibinfo {author} {\bibfnamefont {M.}~\bibnamefont
			{Buchert}}, \bibinfo {author} {\bibfnamefont {R.}~\bibnamefont {Liu}},
		\bibinfo {author} {\bibfnamefont {S.}~\bibnamefont {Dittrich}},\ and\
		\bibinfo {author} {\bibfnamefont {C.}~\bibnamefont {Merz}},\ }\href
	{http://ressourcenfieber.eu/publications/reports/Rare%20earths%20study_Oeko-Institut_Jan%202011.pdf}
		{\bibfield  {journal} {\bibinfo  {journal} {{\"O}ko-Institut eV Darmstadt}\
			}\textbf {\bibinfo {volume} {49}},\ \bibinfo {pages} {30} (\bibinfo {year}
			{2011})}\BibitemShut {NoStop}%
		\bibitem [{\citenamefont {Zhou}\ and\ \citenamefont
			{Fiete}(2020)}]{zhou2020rare}%
		\BibitemOpen
		\bibfield  {author} {\bibinfo {author} {\bibfnamefont {J.}~\bibnamefont
				{Zhou}}\ and\ \bibinfo {author} {\bibfnamefont {G.~A.}\ \bibnamefont
				{Fiete}},\ }\bibfield  {journal} {\bibinfo  {journal} {Physics Today}\
		}\textbf {\bibinfo {volume} {73}},\ \href {https://doi.org/10.1063/PT.3.4397}
		{10.1063/PT.3.4397} (\bibinfo {year} {2020})\BibitemShut {NoStop}%
		\bibitem [{\citenamefont {Scheie}\ \emph {et~al.}(2021)\citenamefont {Scheie},
			\citenamefont {Laurell}, \citenamefont {McClarty}, \citenamefont {Granroth},
			\citenamefont {Stone}, \citenamefont {Moessner},\ and\ \citenamefont
			{Nagler}}]{Scheie_Gd_PRL}%
		\BibitemOpen
		\bibfield  {author} {\bibinfo {author} {\bibfnamefont {A.}~\bibnamefont
				{Scheie}}, \bibinfo {author} {\bibfnamefont {P.}~\bibnamefont {Laurell}},
			\bibinfo {author} {\bibfnamefont {P.~A.}\ \bibnamefont {McClarty}}, \bibinfo
			{author} {\bibfnamefont {G.}~\bibnamefont {Granroth}}, \bibinfo {author}
			{\bibfnamefont {M.~B.}\ \bibnamefont {Stone}}, \bibinfo {author}
			{\bibfnamefont {R.}~\bibnamefont {Moessner}},\ and\ \bibinfo {author}
			{\bibfnamefont {S.~E.}\ \bibnamefont {Nagler}},\ }\href@noop {} {\bibfield
			{journal} {\bibinfo  {journal} {in preparation}\ } (\bibinfo {year}
			{2021})}\BibitemShut {NoStop}%
		\bibitem [{\citenamefont {Nigh}\ \emph {et~al.}(1963)\citenamefont {Nigh},
			\citenamefont {Legvold},\ and\ \citenamefont {Spedding}}]{Nigh_1963}%
		\BibitemOpen
		\bibfield  {author} {\bibinfo {author} {\bibfnamefont {H.~E.}\ \bibnamefont
				{Nigh}}, \bibinfo {author} {\bibfnamefont {S.}~\bibnamefont {Legvold}},\ and\
			\bibinfo {author} {\bibfnamefont {F.~H.}\ \bibnamefont {Spedding}},\ }\href
		{https://doi.org/10.1103/PhysRev.132.1092} {\bibfield  {journal} {\bibinfo
				{journal} {Phys. Rev.}\ }\textbf {\bibinfo {volume} {132}},\ \bibinfo {pages}
			{1092} (\bibinfo {year} {1963})}\BibitemShut {NoStop}%
		\bibitem [{\citenamefont {Cable}\ and\ \citenamefont
			{Wollan}(1968)}]{Cable_1968}%
		\BibitemOpen
		\bibfield  {author} {\bibinfo {author} {\bibfnamefont {J.~W.}\ \bibnamefont
				{Cable}}\ and\ \bibinfo {author} {\bibfnamefont {E.~O.}\ \bibnamefont
				{Wollan}},\ }\href {https://doi.org/10.1103/PhysRev.165.733} {\bibfield
			{journal} {\bibinfo  {journal} {Phys. Rev.}\ }\textbf {\bibinfo {volume}
				{165}},\ \bibinfo {pages} {733} (\bibinfo {year} {1968})}\BibitemShut
		{NoStop}%
		\bibitem [{\citenamefont {Moon}\ \emph {et~al.}(1972)\citenamefont {Moon},
			\citenamefont {Koehler}, \citenamefont {Cable},\ and\ \citenamefont
			{Child}}]{Moon_1972}%
		\BibitemOpen
		\bibfield  {author} {\bibinfo {author} {\bibfnamefont {R.~M.}\ \bibnamefont
				{Moon}}, \bibinfo {author} {\bibfnamefont {W.~C.}\ \bibnamefont {Koehler}},
			\bibinfo {author} {\bibfnamefont {J.~W.}\ \bibnamefont {Cable}},\ and\
			\bibinfo {author} {\bibfnamefont {H.~R.}\ \bibnamefont {Child}},\ }\href
		{https://doi.org/10.1103/PhysRevB.5.997} {\bibfield  {journal} {\bibinfo
				{journal} {Phys. Rev. B}\ }\textbf {\bibinfo {volume} {5}},\ \bibinfo {pages}
			{997} (\bibinfo {year} {1972})}\BibitemShut {NoStop}%
		\bibitem [{\citenamefont {Kip}(1953)}]{Kip_1953}%
		\BibitemOpen
		\bibfield  {author} {\bibinfo {author} {\bibfnamefont {A.~F.}\ \bibnamefont
				{Kip}},\ }\href {https://doi.org/10.1103/RevModPhys.25.229} {\bibfield
			{journal} {\bibinfo  {journal} {Rev. Mod. Phys.}\ }\textbf {\bibinfo {volume}
				{25}},\ \bibinfo {pages} {229} (\bibinfo {year} {1953})}\BibitemShut
		{NoStop}%
		\bibitem [{\citenamefont {Franse}\ and\ \citenamefont
			{Gersdorf}(1980)}]{Franse_1980}%
		\BibitemOpen
		\bibfield  {author} {\bibinfo {author} {\bibfnamefont {J.~J.~M.}\
				\bibnamefont {Franse}}\ and\ \bibinfo {author} {\bibfnamefont
				{R.}~\bibnamefont {Gersdorf}},\ }\href
		{https://doi.org/10.1103/PhysRevLett.45.50} {\bibfield  {journal} {\bibinfo
				{journal} {Phys. Rev. Lett.}\ }\textbf {\bibinfo {volume} {45}},\ \bibinfo
			{pages} {50} (\bibinfo {year} {1980})}\BibitemShut {NoStop}%
		\bibitem [{\citenamefont {Coey}\ \emph {et~al.}(1999)\citenamefont {Coey},
			\citenamefont {Skumryev},\ and\ \citenamefont {Gallagher}}]{Coey1999}%
		\BibitemOpen
		\bibfield  {author} {\bibinfo {author} {\bibfnamefont {J.~M.~D.}\
				\bibnamefont {Coey}}, \bibinfo {author} {\bibfnamefont {V.}~\bibnamefont
				{Skumryev}},\ and\ \bibinfo {author} {\bibfnamefont {K.}~\bibnamefont
				{Gallagher}},\ }\href {https://doi.org/10.1038/43363} {\bibfield  {journal}
			{\bibinfo  {journal} {Nature}\ }\textbf {\bibinfo {volume} {401}},\ \bibinfo
			{pages} {35} (\bibinfo {year} {1999})}\BibitemShut {NoStop}%
		\bibitem [{\citenamefont {Colarieti-Tosti}\ \emph {et~al.}(2005)\citenamefont
			{Colarieti-Tosti}, \citenamefont {Burkert}, \citenamefont {Eriksson},
			\citenamefont {Nordstr\"om},\ and\ \citenamefont {Brooks}}]{Tosti_2005}%
		\BibitemOpen
		\bibfield  {author} {\bibinfo {author} {\bibfnamefont {M.}~\bibnamefont
				{Colarieti-Tosti}}, \bibinfo {author} {\bibfnamefont {T.}~\bibnamefont
				{Burkert}}, \bibinfo {author} {\bibfnamefont {O.}~\bibnamefont {Eriksson}},
			\bibinfo {author} {\bibfnamefont {L.}~\bibnamefont {Nordstr\"om}},\ and\
			\bibinfo {author} {\bibfnamefont {M.~S.~S.}\ \bibnamefont {Brooks}},\ }\href
		{https://doi.org/10.1103/PhysRevB.72.094423} {\bibfield  {journal} {\bibinfo
				{journal} {Phys. Rev. B}\ }\textbf {\bibinfo {volume} {72}},\ \bibinfo
			{pages} {094423} (\bibinfo {year} {2005})}\BibitemShut {NoStop}%
		\bibitem [{\citenamefont {Abdelouahed}\ and\ \citenamefont
			{Alouani}(2009)}]{PhysRevB.79.054406}%
		\BibitemOpen
		\bibfield  {author} {\bibinfo {author} {\bibfnamefont {S.}~\bibnamefont
				{Abdelouahed}}\ and\ \bibinfo {author} {\bibfnamefont {M.}~\bibnamefont
				{Alouani}},\ }\href {https://doi.org/10.1103/PhysRevB.79.054406} {\bibfield
			{journal} {\bibinfo  {journal} {Phys. Rev. B}\ }\textbf {\bibinfo {volume}
				{79}},\ \bibinfo {pages} {054406} (\bibinfo {year} {2009})}\BibitemShut
		{NoStop}%
		\bibitem [{\citenamefont {Koehler}\ \emph {et~al.}(1970)\citenamefont
			{Koehler}, \citenamefont {Child}, \citenamefont {Nicklow}, \citenamefont
			{Smith}, \citenamefont {Moon},\ and\ \citenamefont {Cable}}]{Koehler_1970}%
		\BibitemOpen
		\bibfield  {author} {\bibinfo {author} {\bibfnamefont {W.~C.}\ \bibnamefont
				{Koehler}}, \bibinfo {author} {\bibfnamefont {H.~R.}\ \bibnamefont {Child}},
			\bibinfo {author} {\bibfnamefont {R.~M.}\ \bibnamefont {Nicklow}}, \bibinfo
			{author} {\bibfnamefont {H.~G.}\ \bibnamefont {Smith}}, \bibinfo {author}
			{\bibfnamefont {R.~M.}\ \bibnamefont {Moon}},\ and\ \bibinfo {author}
			{\bibfnamefont {J.~W.}\ \bibnamefont {Cable}},\ }\href
		{https://doi.org/10.1103/PhysRevLett.24.16} {\bibfield  {journal} {\bibinfo
				{journal} {Phys. Rev. Lett.}\ }\textbf {\bibinfo {volume} {24}},\ \bibinfo
			{pages} {16} (\bibinfo {year} {1970})}\BibitemShut {NoStop}%
		\bibitem [{\citenamefont {Squires}(2012)}]{Squires}%
		\BibitemOpen
		\bibfield  {author} {\bibinfo {author} {\bibfnamefont {G.~L.}\ \bibnamefont
				{Squires}},\ }\href@noop {} {\emph {\bibinfo {title} {Introduction to the
					Theory of Thermal Neutron Scattering}}},\ \bibinfo {edition} {3rd}\ ed.\
		(\bibinfo  {publisher} {Cambridge University Press},\ \bibinfo {address}
		{Cambridge, UK},\ \bibinfo {year} {2012})\BibitemShut {NoStop}%
		\bibitem [{\citenamefont {Jensen}\ and\ \citenamefont
			{Mackintosh}(1991)}]{Jensen+Mackintosh}%
		\BibitemOpen
		\bibfield  {author} {\bibinfo {author} {\bibfnamefont {J.}~\bibnamefont
				{Jensen}}\ and\ \bibinfo {author} {\bibfnamefont {A.}~\bibnamefont
				{Mackintosh}},\ }\href@noop {} {\emph {\bibinfo {title} {Rare Earth
					Magnetism, Structures and Excitations}}}\ (\bibinfo  {publisher} {Clarendon
			Press},\ \bibinfo {address} {Oxford, UK},\ \bibinfo {year}
		{1991})\BibitemShut {NoStop}%
		\bibitem [{\citenamefont {Lindg\aa{}rd}(1978)}]{Lindgard_1978}%
		\BibitemOpen
		\bibfield  {author} {\bibinfo {author} {\bibfnamefont {P.-A.}\ \bibnamefont
				{Lindg\aa{}rd}},\ }\href {https://doi.org/10.1103/PhysRevB.17.2348}
		{\bibfield  {journal} {\bibinfo  {journal} {Phys. Rev. B}\ }\textbf {\bibinfo
				{volume} {17}},\ \bibinfo {pages} {2348} (\bibinfo {year}
			{1978})}\BibitemShut {NoStop}%
		\bibitem [{\citenamefont {Stone}\ \emph {et~al.}(2014)\citenamefont {Stone},
			\citenamefont {Niedziela}, \citenamefont {Abernathy}, \citenamefont
			{DeBeer-Schmitt}, \citenamefont {Ehlers}, \citenamefont {Garlea},
			\citenamefont {Granroth}, \citenamefont {Graves-Brook}, \citenamefont
			{Kolesnikov}, \citenamefont {Podlesnyak},\ and\ \citenamefont
			{Winn}}]{stone2014comparison}%
		\BibitemOpen
		\bibfield  {author} {\bibinfo {author} {\bibfnamefont {M.~B.}\ \bibnamefont
				{Stone}}, \bibinfo {author} {\bibfnamefont {J.~L.}\ \bibnamefont
				{Niedziela}}, \bibinfo {author} {\bibfnamefont {D.~L.}\ \bibnamefont
				{Abernathy}}, \bibinfo {author} {\bibfnamefont {L.}~\bibnamefont
				{DeBeer-Schmitt}}, \bibinfo {author} {\bibfnamefont {G.}~\bibnamefont
				{Ehlers}}, \bibinfo {author} {\bibfnamefont {O.}~\bibnamefont {Garlea}},
			\bibinfo {author} {\bibfnamefont {G.~E.}\ \bibnamefont {Granroth}}, \bibinfo
			{author} {\bibfnamefont {M.}~\bibnamefont {Graves-Brook}}, \bibinfo {author}
			{\bibfnamefont {A.~I.}\ \bibnamefont {Kolesnikov}}, \bibinfo {author}
			{\bibfnamefont {A.}~\bibnamefont {Podlesnyak}},\ and\ \bibinfo {author}
			{\bibfnamefont {B.}~\bibnamefont {Winn}},\ }\href
		{https://doi.org/10.1063/1.4870050} {\bibfield  {journal} {\bibinfo
				{journal} {Review of Scientific Instruments}\ }\textbf {\bibinfo {volume}
				{85}},\ \bibinfo {pages} {045113} (\bibinfo {year} {2014})}\BibitemShut
		{NoStop}%
		\bibitem [{\citenamefont {Granroth}\ \emph {et~al.}(2010)\citenamefont
			{Granroth}, \citenamefont {Kolesnikov}, \citenamefont {Sherline},
			\citenamefont {Clancy}, \citenamefont {Ross}, \citenamefont {Ruff},
			\citenamefont {Gaulin},\ and\ \citenamefont {Nagler}}]{Granroth2010}%
		\BibitemOpen
		\bibfield  {author} {\bibinfo {author} {\bibfnamefont {G.~E.}\ \bibnamefont
				{Granroth}}, \bibinfo {author} {\bibfnamefont {A.~I.}\ \bibnamefont
				{Kolesnikov}}, \bibinfo {author} {\bibfnamefont {T.~E.}\ \bibnamefont
				{Sherline}}, \bibinfo {author} {\bibfnamefont {J.~P.}\ \bibnamefont
				{Clancy}}, \bibinfo {author} {\bibfnamefont {K.~A.}\ \bibnamefont {Ross}},
			\bibinfo {author} {\bibfnamefont {J.~P.~C.}\ \bibnamefont {Ruff}}, \bibinfo
			{author} {\bibfnamefont {B.~D.}\ \bibnamefont {Gaulin}},\ and\ \bibinfo
			{author} {\bibfnamefont {S.~E.}\ \bibnamefont {Nagler}},\ }\href
		{http://stacks.iop.org/1742-6596/251/i=1/a=012058} {\bibfield  {journal}
			{\bibinfo  {journal} {Journal of Physics: Conference Series}\ }\textbf
			{\bibinfo {volume} {251}},\ \bibinfo {pages} {012058} (\bibinfo {year}
			{2010})}\BibitemShut {NoStop}%
		\bibitem [{\citenamefont {Granroth}\ \emph {et~al.}(2006)\citenamefont
			{Granroth}, \citenamefont {Vandergriff},\ and\ \citenamefont
			{Nagler}}]{Granroth2006}%
		\BibitemOpen
		\bibfield  {author} {\bibinfo {author} {\bibfnamefont {G.~E.}\ \bibnamefont
				{Granroth}}, \bibinfo {author} {\bibfnamefont {D.~H.}\ \bibnamefont
				{Vandergriff}},\ and\ \bibinfo {author} {\bibfnamefont {S.~E.}\ \bibnamefont
				{Nagler}},\ }\href {https://doi.org/10.1016/j.physb.2006.05.379} {\bibfield
			{journal} {\bibinfo  {journal} {Physica B: Condensed Matter}\ }\textbf
			{\bibinfo {volume} {385-86}},\ \bibinfo {pages} {1104} (\bibinfo {year}
			{2006})}\BibitemShut {NoStop}%
		\bibitem [{\citenamefont {Mason}\ \emph {et~al.}(2006)\citenamefont {Mason},
			\citenamefont {Abernathy}, \citenamefont {Anderson}, \citenamefont {Ankner},
			\citenamefont {Egami}, \citenamefont {Ehlers}, \citenamefont {Ekkebus},
			\citenamefont {Granroth}, \citenamefont {Hagen}, \citenamefont {Herwig},
			\citenamefont {Hodges}, \citenamefont {Hoffmann}, \citenamefont {Horak},
			\citenamefont {Horton}, \citenamefont {Klose}, \citenamefont {Larese},
			\citenamefont {Mesecar}, \citenamefont {Myles}, \citenamefont {Neuefeind},
			\citenamefont {Ohl}, \citenamefont {Tulk}, \citenamefont {Wang},\ and\
			\citenamefont {Zhao}}]{mason2006spallation}%
		\BibitemOpen
		\bibfield  {author} {\bibinfo {author} {\bibfnamefont {T.~E.}\ \bibnamefont
				{Mason}}, \bibinfo {author} {\bibfnamefont {D.}~\bibnamefont {Abernathy}},
			\bibinfo {author} {\bibfnamefont {I.}~\bibnamefont {Anderson}}, \bibinfo
			{author} {\bibfnamefont {J.}~\bibnamefont {Ankner}}, \bibinfo {author}
			{\bibfnamefont {T.}~\bibnamefont {Egami}}, \bibinfo {author} {\bibfnamefont
				{G.}~\bibnamefont {Ehlers}}, \bibinfo {author} {\bibfnamefont
				{A.}~\bibnamefont {Ekkebus}}, \bibinfo {author} {\bibfnamefont
				{G.}~\bibnamefont {Granroth}}, \bibinfo {author} {\bibfnamefont
				{M.}~\bibnamefont {Hagen}}, \bibinfo {author} {\bibfnamefont
				{K.}~\bibnamefont {Herwig}}, \bibinfo {author} {\bibfnamefont
				{J.}~\bibnamefont {Hodges}}, \bibinfo {author} {\bibfnamefont
				{C.}~\bibnamefont {Hoffmann}}, \bibinfo {author} {\bibfnamefont
				{C.}~\bibnamefont {Horak}}, \bibinfo {author} {\bibfnamefont
				{L.}~\bibnamefont {Horton}}, \bibinfo {author} {\bibfnamefont
				{F.}~\bibnamefont {Klose}}, \bibinfo {author} {\bibfnamefont
				{J.}~\bibnamefont {Larese}}, \bibinfo {author} {\bibfnamefont
				{A.}~\bibnamefont {Mesecar}}, \bibinfo {author} {\bibfnamefont
				{D.}~\bibnamefont {Myles}}, \bibinfo {author} {\bibfnamefont
				{J.}~\bibnamefont {Neuefeind}}, \bibinfo {author} {\bibfnamefont
				{M.}~\bibnamefont {Ohl}}, \bibinfo {author} {\bibfnamefont {C.}~\bibnamefont
				{Tulk}}, \bibinfo {author} {\bibfnamefont {X.-L.}\ \bibnamefont {Wang}},\
			and\ \bibinfo {author} {\bibfnamefont {J.}~\bibnamefont {Zhao}},\ }\href
		{https://doi.org/10.1016/j.physb.2006.05.281} {\bibfield  {journal} {\bibinfo
				{journal} {Physica B: Condensed Matter}\ }\textbf {\bibinfo {volume}
				{385}},\ \bibinfo {pages} {955} (\bibinfo {year} {2006})}\BibitemShut
		{NoStop}%
		\bibitem [{\citenamefont {Arnold}\ \emph {et~al.}(2014)\citenamefont {Arnold},
			\citenamefont {Bilheux}, \citenamefont {Borreguero}, \citenamefont {Buts},
			\citenamefont {Campbell}, \citenamefont {Chapon}, \citenamefont {Doucet},
			\citenamefont {Draper}, \citenamefont {{Ferraz Leal}}, \citenamefont {Gigg},
			\citenamefont {Lynch}, \citenamefont {Markvardsen}, \citenamefont
			{Mikkelson}, \citenamefont {Mikkelson}, \citenamefont {Miller}, \citenamefont
			{Palmen}, \citenamefont {Parker}, \citenamefont {Passos}, \citenamefont
			{Perring}, \citenamefont {Peterson}, \citenamefont {Ren}, \citenamefont
			{Reuter}, \citenamefont {Savici}, \citenamefont {Taylor}, \citenamefont
			{Taylor}, \citenamefont {Tolchenov}, \citenamefont {Zhou},\ and\
			\citenamefont {Zikovsky}}]{Mantid2014}%
		\BibitemOpen
		\bibfield  {author} {\bibinfo {author} {\bibfnamefont {O.}~\bibnamefont
				{Arnold}}, \bibinfo {author} {\bibfnamefont {J.}~\bibnamefont {Bilheux}},
			\bibinfo {author} {\bibfnamefont {J.}~\bibnamefont {Borreguero}}, \bibinfo
			{author} {\bibfnamefont {A.}~\bibnamefont {Buts}}, \bibinfo {author}
			{\bibfnamefont {S.}~\bibnamefont {Campbell}}, \bibinfo {author}
			{\bibfnamefont {L.}~\bibnamefont {Chapon}}, \bibinfo {author} {\bibfnamefont
				{M.}~\bibnamefont {Doucet}}, \bibinfo {author} {\bibfnamefont
				{N.}~\bibnamefont {Draper}}, \bibinfo {author} {\bibfnamefont
				{R.}~\bibnamefont {{Ferraz Leal}}}, \bibinfo {author} {\bibfnamefont
				{M.}~\bibnamefont {Gigg}}, \bibinfo {author} {\bibfnamefont {V.}~\bibnamefont
				{Lynch}}, \bibinfo {author} {\bibfnamefont {A.}~\bibnamefont {Markvardsen}},
			\bibinfo {author} {\bibfnamefont {D.}~\bibnamefont {Mikkelson}}, \bibinfo
			{author} {\bibfnamefont {R.}~\bibnamefont {Mikkelson}}, \bibinfo {author}
			{\bibfnamefont {R.}~\bibnamefont {Miller}}, \bibinfo {author} {\bibfnamefont
				{K.}~\bibnamefont {Palmen}}, \bibinfo {author} {\bibfnamefont
				{P.}~\bibnamefont {Parker}}, \bibinfo {author} {\bibfnamefont
				{G.}~\bibnamefont {Passos}}, \bibinfo {author} {\bibfnamefont
				{T.}~\bibnamefont {Perring}}, \bibinfo {author} {\bibfnamefont
				{P.}~\bibnamefont {Peterson}}, \bibinfo {author} {\bibfnamefont
				{S.}~\bibnamefont {Ren}}, \bibinfo {author} {\bibfnamefont {M.}~\bibnamefont
				{Reuter}}, \bibinfo {author} {\bibfnamefont {A.}~\bibnamefont {Savici}},
			\bibinfo {author} {\bibfnamefont {J.}~\bibnamefont {Taylor}}, \bibinfo
			{author} {\bibfnamefont {R.}~\bibnamefont {Taylor}}, \bibinfo {author}
			{\bibfnamefont {R.}~\bibnamefont {Tolchenov}}, \bibinfo {author}
			{\bibfnamefont {W.}~\bibnamefont {Zhou}},\ and\ \bibinfo {author}
			{\bibfnamefont {J.}~\bibnamefont {Zikovsky}},\ }\href
		{https://doi.org/https://doi.org/10.1016/j.nima.2014.07.029} {\bibfield
			{journal} {\bibinfo  {journal} {Nuclear Instruments and Methods in Physics
					Research Section A: Accelerators, Spectrometers, Detectors and Associated
					Equipment}\ }\textbf {\bibinfo {volume} {764}},\ \bibinfo {pages} {156}
			(\bibinfo {year} {2014})}\BibitemShut {NoStop}%
		\bibitem [{Sup()}]{SuppMat}%
		\BibitemOpen
		\href@noop {} {}\bibinfo {note} {See Supplemental Material at [URL will be
			inserted by publisher] for more details of the experiments and
			calculations.}\BibitemShut {Stop}%
		\bibitem [{\citenamefont {Virtanen}\ \emph {et~al.}(2020)\citenamefont
			{Virtanen}, \citenamefont {Gommers}, \citenamefont {Oliphant}, \citenamefont
			{Haberland}, \citenamefont {Reddy}, \citenamefont {Cournapeau}, \citenamefont
			{Burovski}, \citenamefont {Peterson}, \citenamefont {Weckesser},
			\citenamefont {Bright} \emph {et~al.}}]{virtanen2020scipy}%
		\BibitemOpen
		\bibfield  {author} {\bibinfo {author} {\bibfnamefont {P.}~\bibnamefont
				{Virtanen}}, \bibinfo {author} {\bibfnamefont {R.}~\bibnamefont {Gommers}},
			\bibinfo {author} {\bibfnamefont {T.~E.}\ \bibnamefont {Oliphant}}, \bibinfo
			{author} {\bibfnamefont {M.}~\bibnamefont {Haberland}}, \bibinfo {author}
			{\bibfnamefont {T.}~\bibnamefont {Reddy}}, \bibinfo {author} {\bibfnamefont
				{D.}~\bibnamefont {Cournapeau}}, \bibinfo {author} {\bibfnamefont
				{E.}~\bibnamefont {Burovski}}, \bibinfo {author} {\bibfnamefont
				{P.}~\bibnamefont {Peterson}}, \bibinfo {author} {\bibfnamefont
				{W.}~\bibnamefont {Weckesser}}, \bibinfo {author} {\bibfnamefont
				{J.}~\bibnamefont {Bright}}, \emph {et~al.},\ }\href
		{https://doi.org/10.1038/s41592-019-0686-2} {\bibfield  {journal} {\bibinfo
				{journal} {Nature Methods}\ }\textbf {\bibinfo {volume} {17}},\ \bibinfo
			{pages} {261} (\bibinfo {year} {2020})}\BibitemShut {NoStop}%
		\bibitem [{\citenamefont {Cable}\ \emph {et~al.}(1981)\citenamefont {Cable},
			\citenamefont {Wakabayashi},\ and\ \citenamefont {Nicklow}}]{Cable_1981}%
		\BibitemOpen
		\bibfield  {author} {\bibinfo {author} {\bibfnamefont {J.~W.}\ \bibnamefont
				{Cable}}, \bibinfo {author} {\bibfnamefont {N.}~\bibnamefont {Wakabayashi}},\
			and\ \bibinfo {author} {\bibfnamefont {R.~M.}\ \bibnamefont {Nicklow}},\
		}\href {https://doi.org/10.1063/1.328888} {\bibfield  {journal} {\bibinfo
				{journal} {Journal of Applied Physics}\ }\textbf {\bibinfo {volume} {52}},\
			\bibinfo {pages} {2231} (\bibinfo {year} {1981})}\BibitemShut {NoStop}%
		\bibitem [{\citenamefont {Cable}\ \emph {et~al.}(1985)\citenamefont {Cable},
			\citenamefont {Nicklow},\ and\ \citenamefont {Wakabayashi}}]{Cable_1985}%
		\BibitemOpen
		\bibfield  {author} {\bibinfo {author} {\bibfnamefont {J.~W.}\ \bibnamefont
				{Cable}}, \bibinfo {author} {\bibfnamefont {R.~M.}\ \bibnamefont {Nicklow}},\
			and\ \bibinfo {author} {\bibfnamefont {N.}~\bibnamefont {Wakabayashi}},\
		}\href {https://doi.org/10.1103/PhysRevB.32.1710} {\bibfield  {journal}
			{\bibinfo  {journal} {Phys. Rev. B}\ }\textbf {\bibinfo {volume} {32}},\
			\bibinfo {pages} {1710} (\bibinfo {year} {1985})}\BibitemShut {NoStop}%
		\bibitem [{\citenamefont {Cable}\ and\ \citenamefont
			{Nicklow}(1989)}]{Cable_1989}%
		\BibitemOpen
		\bibfield  {author} {\bibinfo {author} {\bibfnamefont {J.~W.}\ \bibnamefont
				{Cable}}\ and\ \bibinfo {author} {\bibfnamefont {R.~M.}\ \bibnamefont
				{Nicklow}},\ }\href {https://doi.org/10.1103/PhysRevB.39.11732} {\bibfield
			{journal} {\bibinfo  {journal} {Phys. Rev. B}\ }\textbf {\bibinfo {volume}
				{39}},\ \bibinfo {pages} {11732} (\bibinfo {year} {1989})}\BibitemShut
		{NoStop}%
		\bibitem [{\citenamefont {Ruderman}\ and\ \citenamefont
			{Kittel}(1954)}]{RK_1954}%
		\BibitemOpen
		\bibfield  {author} {\bibinfo {author} {\bibfnamefont {M.~A.}\ \bibnamefont
				{Ruderman}}\ and\ \bibinfo {author} {\bibfnamefont {C.}~\bibnamefont
				{Kittel}},\ }\href {https://doi.org/10.1103/PhysRev.96.99} {\bibfield
			{journal} {\bibinfo  {journal} {Phys. Rev.}\ }\textbf {\bibinfo {volume}
				{96}},\ \bibinfo {pages} {99} (\bibinfo {year} {1954})}\BibitemShut {NoStop}%
		\bibitem [{\citenamefont {Kasuya}(1956)}]{Kasuya_1956}%
		\BibitemOpen
		\bibfield  {author} {\bibinfo {author} {\bibfnamefont {T.}~\bibnamefont
				{Kasuya}},\ }\href {https://doi.org/10.1143/PTP.16.45} {\bibfield  {journal}
			{\bibinfo  {journal} {Progress of Theoretical Physics}\ }\textbf {\bibinfo
				{volume} {16}},\ \bibinfo {pages} {45} (\bibinfo {year} {1956})}\BibitemShut
		{NoStop}%
		\bibitem [{\citenamefont {Yosida}(1957)}]{Yosida_1957}%
		\BibitemOpen
		\bibfield  {author} {\bibinfo {author} {\bibfnamefont {K.}~\bibnamefont
				{Yosida}},\ }\href {https://doi.org/10.1103/PhysRev.106.893} {\bibfield
			{journal} {\bibinfo  {journal} {Phys. Rev.}\ }\textbf {\bibinfo {volume}
				{106}},\ \bibinfo {pages} {893} (\bibinfo {year} {1957})}\BibitemShut
		{NoStop}%
		\bibitem [{\citenamefont {Mattocks}\ and\ \citenamefont
			{Young}(1977)}]{Mattocks_1977}%
		\BibitemOpen
		\bibfield  {author} {\bibinfo {author} {\bibfnamefont {P.~G.}\ \bibnamefont
				{Mattocks}}\ and\ \bibinfo {author} {\bibfnamefont {R.~C.}\ \bibnamefont
				{Young}},\ }\href {https://doi.org/10.1088/0305-4608/7/7/021} {\bibfield
			{journal} {\bibinfo  {journal} {Journal of Physics F: Metal Physics}\
			}\textbf {\bibinfo {volume} {7}},\ \bibinfo {pages} {1219} (\bibinfo {year}
			{1977})}\BibitemShut {NoStop}%
		\bibitem [{\citenamefont {Turek}\ \emph {et~al.}(2003)\citenamefont {Turek},
			\citenamefont {Kudrnovsk}, \citenamefont {Bihlmayer},\ and\ \citenamefont
			{Bl\"ugel}}]{Turek_2003}%
		\BibitemOpen
		\bibfield  {author} {\bibinfo {author} {\bibfnamefont {I.}~\bibnamefont
				{Turek}}, \bibinfo {author} {\bibfnamefont {J.}~\bibnamefont {Kudrnovsk}},
			\bibinfo {author} {\bibfnamefont {G.}~\bibnamefont {Bihlmayer}},\ and\
			\bibinfo {author} {\bibfnamefont {S.}~\bibnamefont {Bl\"ugel}},\ }\href
		{https://doi.org/10.1088/0953-8984/15/17/327} {\bibfield  {journal} {\bibinfo
				{journal} {Journal of Physics: Condensed Matter}\ }\textbf {\bibinfo
				{volume} {15}},\ \bibinfo {pages} {2771} (\bibinfo {year}
			{2003})}\BibitemShut {NoStop}%
		\bibitem [{\citenamefont {Hindmarch}\ and\ \citenamefont
			{Hickey}(2003)}]{Hindmarch_2003}%
		\BibitemOpen
		\bibfield  {author} {\bibinfo {author} {\bibfnamefont {A.~T.}\ \bibnamefont
				{Hindmarch}}\ and\ \bibinfo {author} {\bibfnamefont {B.~J.}\ \bibnamefont
				{Hickey}},\ }\href {https://doi.org/10.1103/PhysRevLett.91.116601} {\bibfield
			{journal} {\bibinfo  {journal} {Phys. Rev. Lett.}\ }\textbf {\bibinfo
				{volume} {91}},\ \bibinfo {pages} {116601} (\bibinfo {year}
			{2003})}\BibitemShut {NoStop}%
		\bibitem [{\citenamefont {Watson}\ and\ \citenamefont
			{Freeman}(1969)}]{Watson_1969}%
		\BibitemOpen
		\bibfield  {author} {\bibinfo {author} {\bibfnamefont {R.~E.}\ \bibnamefont
				{Watson}}\ and\ \bibinfo {author} {\bibfnamefont {A.~J.}\ \bibnamefont
				{Freeman}},\ }\href {https://doi.org/10.1103/PhysRev.178.725} {\bibfield
			{journal} {\bibinfo  {journal} {Phys. Rev.}\ }\textbf {\bibinfo {volume}
				{178}},\ \bibinfo {pages} {725} (\bibinfo {year} {1969})}\BibitemShut
		{NoStop}%
		\bibitem [{\citenamefont {Lindg\aa{}rd}\ \emph {et~al.}(1975)\citenamefont
			{Lindg\aa{}rd}, \citenamefont {Harmon},\ and\ \citenamefont
			{Freeman}}]{Lindgard_1975}%
		\BibitemOpen
		\bibfield  {author} {\bibinfo {author} {\bibfnamefont {P.~A.}\ \bibnamefont
				{Lindg\aa{}rd}}, \bibinfo {author} {\bibfnamefont {B.~N.}\ \bibnamefont
				{Harmon}},\ and\ \bibinfo {author} {\bibfnamefont {A.~J.}\ \bibnamefont
				{Freeman}},\ }\href {https://doi.org/10.1103/PhysRevLett.35.383} {\bibfield
			{journal} {\bibinfo  {journal} {Phys. Rev. Lett.}\ }\textbf {\bibinfo
				{volume} {35}},\ \bibinfo {pages} {383} (\bibinfo {year} {1975})}\BibitemShut
		{NoStop}%
		\bibitem [{\citenamefont {Brinkman}(1967)}]{Brinkman_1967}%
		\BibitemOpen
		\bibfield  {author} {\bibinfo {author} {\bibfnamefont {W.}~\bibnamefont
				{Brinkman}},\ }\href {https://doi.org/10.1063/1.1709692} {\bibfield
			{journal} {\bibinfo  {journal} {Journal of Applied Physics}\ }\textbf
			{\bibinfo {volume} {38}},\ \bibinfo {pages} {939} (\bibinfo {year}
			{1967})}\BibitemShut {NoStop}%
		\bibitem [{\citenamefont {Cracknell}(1970)}]{Cracknell_1970}%
		\BibitemOpen
		\bibfield  {author} {\bibinfo {author} {\bibfnamefont {A.~P.}\ \bibnamefont
				{Cracknell}},\ }\href {https://doi.org/10.1088/0022-3719/3/2s/308} {\bibfield
			{journal} {\bibinfo  {journal} {J. Phys. C: Solid State Phys.}\ }\textbf
			{\bibinfo {volume} {3}},\ \bibinfo {pages} {S175} (\bibinfo {year}
			{1970})}\BibitemShut {NoStop}%
		\bibitem [{\citenamefont {McClarty}(2021)}]{McClarty_2021}%
		\BibitemOpen
		\bibfield  {author} {\bibinfo {author} {\bibfnamefont {P.}~\bibnamefont
				{McClarty}},\ }\href {https://arxiv.org/abs/2106.01430} {\bibfield  {journal}
			{\bibinfo  {journal} {arXiv preprint arXiv:2106.01430}\ } (\bibinfo {year}
			{2021})}\BibitemShut {NoStop}%
		\bibitem [{\citenamefont {McClarty}\ and\ \citenamefont
			{Rau}(2019)}]{McClarty_2019}%
		\BibitemOpen
		\bibfield  {author} {\bibinfo {author} {\bibfnamefont {P.~A.}\ \bibnamefont
				{McClarty}}\ and\ \bibinfo {author} {\bibfnamefont {J.~G.}\ \bibnamefont
				{Rau}},\ }\href {https://doi.org/10.1103/PhysRevB.100.100405} {\bibfield
			{journal} {\bibinfo  {journal} {Phys. Rev. B}\ }\textbf {\bibinfo {volume}
				{100}},\ \bibinfo {pages} {100405} (\bibinfo {year} {2019})}\BibitemShut
		{NoStop}%
		\bibitem [{\citenamefont {Shivam}\ \emph {et~al.}(2017)\citenamefont {Shivam},
			\citenamefont {Coldea}, \citenamefont {Moessner},\ and\ \citenamefont
			{McClarty}}]{shivam2017neutron}%
		\BibitemOpen
		\bibfield  {author} {\bibinfo {author} {\bibfnamefont {S.}~\bibnamefont
				{Shivam}}, \bibinfo {author} {\bibfnamefont {R.}~\bibnamefont {Coldea}},
			\bibinfo {author} {\bibfnamefont {R.}~\bibnamefont {Moessner}},\ and\
			\bibinfo {author} {\bibfnamefont {P.}~\bibnamefont {McClarty}},\ }\href
		{https://arxiv.org/abs/1712.08535} {\bibfield  {journal} {\bibinfo  {journal}
				{arXiv preprint arXiv:1712.08535}\ } (\bibinfo {year} {2017})}\BibitemShut
		{NoStop}%
		\bibitem [{\citenamefont {Fujiki}\ \emph {et~al.}(1987)\citenamefont {Fujiki},
			\citenamefont {De'Bell},\ and\ \citenamefont {Geldart}}]{Fujiki_1987}%
		\BibitemOpen
		\bibfield  {author} {\bibinfo {author} {\bibfnamefont {N.~M.}\ \bibnamefont
				{Fujiki}}, \bibinfo {author} {\bibfnamefont {K.}~\bibnamefont {De'Bell}},\
			and\ \bibinfo {author} {\bibfnamefont {D.~J.~W.}\ \bibnamefont {Geldart}},\
		}\href {https://doi.org/10.1103/PhysRevB.36.8512} {\bibfield  {journal}
			{\bibinfo  {journal} {Phys. Rev. B}\ }\textbf {\bibinfo {volume} {36}},\
			\bibinfo {pages} {8512} (\bibinfo {year} {1987})}\BibitemShut {NoStop}%
		\bibitem [{\citenamefont {Moriya}(1960)}]{Moriya_1960}%
		\BibitemOpen
		\bibfield  {author} {\bibinfo {author} {\bibfnamefont {T.}~\bibnamefont
				{Moriya}},\ }\href {https://doi.org/10.1103/PhysRev.120.91} {\bibfield
			{journal} {\bibinfo  {journal} {Phys. Rev.}\ }\textbf {\bibinfo {volume}
				{120}},\ \bibinfo {pages} {91} (\bibinfo {year} {1960})}\BibitemShut
		{NoStop}%
		\bibitem [{\citenamefont {Toth}\ and\ \citenamefont {Lake}(2015)}]{SpinW}%
		\BibitemOpen
		\bibfield  {author} {\bibinfo {author} {\bibfnamefont {S.}~\bibnamefont
				{Toth}}\ and\ \bibinfo {author} {\bibfnamefont {B.}~\bibnamefont {Lake}},\
		}\href {https://doi.org/10.1088/0953-8984/27/16/166002} {\bibfield  {journal}
			{\bibinfo  {journal} {Journal of Physics: Condensed Matter}\ }\textbf
			{\bibinfo {volume} {27}},\ \bibinfo {pages} {166002} (\bibinfo {year}
			{2015})}\BibitemShut {NoStop}%
		\bibitem [{\citenamefont {Nicklow}\ \emph {et~al.}(1971)\citenamefont
			{Nicklow}, \citenamefont {Wakabayashi}, \citenamefont {Wilkinson},\ and\
			\citenamefont {Reed}}]{Nicklow_1971}%
		\BibitemOpen
		\bibfield  {author} {\bibinfo {author} {\bibfnamefont {R.~M.}\ \bibnamefont
				{Nicklow}}, \bibinfo {author} {\bibfnamefont {N.}~\bibnamefont
				{Wakabayashi}}, \bibinfo {author} {\bibfnamefont {M.~K.}\ \bibnamefont
				{Wilkinson}},\ and\ \bibinfo {author} {\bibfnamefont {R.~E.}\ \bibnamefont
				{Reed}},\ }\href {https://doi.org/10.1103/PhysRevLett.26.140} {\bibfield
			{journal} {\bibinfo  {journal} {Phys. Rev. Lett.}\ }\textbf {\bibinfo
				{volume} {26}},\ \bibinfo {pages} {140} (\bibinfo {year} {1971})}\BibitemShut
		{NoStop}%
		\bibitem [{\citenamefont {M\o{}ller}\ \emph {et~al.}(1968)\citenamefont
			{M\o{}ller}, \citenamefont {Houmann},\ and\ \citenamefont
			{Mackintosh}}]{Moller_1968}%
		\BibitemOpen
		\bibfield  {author} {\bibinfo {author} {\bibfnamefont {H.~B.}\ \bibnamefont
				{M\o{}ller}}, \bibinfo {author} {\bibfnamefont {J.~C.~G.}\ \bibnamefont
				{Houmann}},\ and\ \bibinfo {author} {\bibfnamefont {A.~R.}\ \bibnamefont
				{Mackintosh}},\ }\href {https://doi.org/10.1063/1.2163623} {\bibfield
			{journal} {\bibinfo  {journal} {Journal of Applied Physics}\ }\textbf
			{\bibinfo {volume} {39}},\ \bibinfo {pages} {807} (\bibinfo {year}
			{1968})}\BibitemShut {NoStop}%
		\bibitem [{\citenamefont {Perring}\ \emph {et~al.}(1995)\citenamefont
			{Perring}, \citenamefont {Taylor},\ and\ \citenamefont
			{Squires}}]{perring1995high}%
		\BibitemOpen
		\bibfield  {author} {\bibinfo {author} {\bibfnamefont {T.}~\bibnamefont
				{Perring}}, \bibinfo {author} {\bibfnamefont {A.}~\bibnamefont {Taylor}},\
			and\ \bibinfo {author} {\bibfnamefont {G.}~\bibnamefont {Squires}},\ }\href
		{https://doi.org/10.1016/0921-4526(95)92829-Y} {\bibfield  {journal}
			{\bibinfo  {journal} {Physica B: Condensed Matter}\ }\textbf {\bibinfo
				{volume} {213}},\ \bibinfo {pages} {348} (\bibinfo {year}
			{1995})}\BibitemShut {NoStop}%
		\bibitem [{\citenamefont {Dietrich}\ and\ \citenamefont
			{Als-Nielsen}(1967)}]{Dietrich_1967}%
		\BibitemOpen
		\bibfield  {author} {\bibinfo {author} {\bibfnamefont {O.~W.}\ \bibnamefont
				{Dietrich}}\ and\ \bibinfo {author} {\bibfnamefont {J.}~\bibnamefont
				{Als-Nielsen}},\ }\href {https://doi.org/10.1103/PhysRev.162.315} {\bibfield
			{journal} {\bibinfo  {journal} {Phys. Rev.}\ }\textbf {\bibinfo {volume}
				{162}},\ \bibinfo {pages} {315} (\bibinfo {year} {1967})}\BibitemShut
		{NoStop}%
		\bibitem [{\citenamefont {Roeland}\ and\ \citenamefont
			{Cock}(1975)}]{Roeland_1975}%
		\BibitemOpen
		\bibfield  {author} {\bibinfo {author} {\bibfnamefont {L.~W.}\ \bibnamefont
				{Roeland}}\ and\ \bibinfo {author} {\bibfnamefont {G.~J.}\ \bibnamefont
				{Cock}},\ }\href {https://doi.org/10.1088/0022-3719/8/20/020} {\bibfield
			{journal} {\bibinfo  {journal} {Journal of Physics C: Solid State Physics}\
			}\textbf {\bibinfo {volume} {8}},\ \bibinfo {pages} {3427} (\bibinfo {year}
			{1975})}\BibitemShut {NoStop}%
		\bibitem [{\citenamefont {Powell}(1964)}]{PowellsMethod}%
		\BibitemOpen
		\bibfield  {author} {\bibinfo {author} {\bibfnamefont {M.~J.~D.}\
				\bibnamefont {Powell}},\ }\href {https://doi.org/10.1093/comjnl/7.2.155}
		{\bibfield  {journal} {\bibinfo  {journal} {The Computer Journal}\ }\textbf
			{\bibinfo {volume} {7}},\ \bibinfo {pages} {155} (\bibinfo {year}
			{1964})}\BibitemShut {NoStop}%
		\bibitem [{\citenamefont {Press}\ \emph {et~al.}(2007)\citenamefont {Press},
			\citenamefont {Teukolsky}, \citenamefont {Vetterling},\ and\ \citenamefont
			{Flannery}}]{NumericalRecipes}%
		\BibitemOpen
		\bibfield  {author} {\bibinfo {author} {\bibfnamefont {W.~H.}\ \bibnamefont
				{Press}}, \bibinfo {author} {\bibfnamefont {S.~A.}\ \bibnamefont
				{Teukolsky}}, \bibinfo {author} {\bibfnamefont {W.~T.}\ \bibnamefont
				{Vetterling}},\ and\ \bibinfo {author} {\bibfnamefont {B.~P.}\ \bibnamefont
				{Flannery}},\ }\href@noop {} {\emph {\bibinfo {title} {Numerical recipes 3rd
					edition: The art of scientific computing}}}\ (\bibinfo  {publisher}
		{Cambridge university press},\ \bibinfo {year} {2007})\BibitemShut {NoStop}%
		\bibitem [{\citenamefont {Pedregosa}\ \emph {et~al.}(2011)\citenamefont
			{Pedregosa}, \citenamefont {Varoquaux}, \citenamefont {Gramfort},
			\citenamefont {Michel}, \citenamefont {Thirion}, \citenamefont {Grisel},
			\citenamefont {Blondel}, \citenamefont {Prettenhofer}, \citenamefont {Weiss},
			\citenamefont {Dubourg}, \citenamefont {Vanderplas}, \citenamefont {Passos},
			\citenamefont {Cournapeau}, \citenamefont {Brucher}, \citenamefont {Perrot},\
			and\ \citenamefont {{{\'E}}douard Duchesnay}}]{scikit}%
		\BibitemOpen
		\bibfield  {author} {\bibinfo {author} {\bibfnamefont {F.}~\bibnamefont
				{Pedregosa}}, \bibinfo {author} {\bibfnamefont {G.}~\bibnamefont
				{Varoquaux}}, \bibinfo {author} {\bibfnamefont {A.}~\bibnamefont {Gramfort}},
			\bibinfo {author} {\bibfnamefont {V.}~\bibnamefont {Michel}}, \bibinfo
			{author} {\bibfnamefont {B.}~\bibnamefont {Thirion}}, \bibinfo {author}
			{\bibfnamefont {O.}~\bibnamefont {Grisel}}, \bibinfo {author} {\bibfnamefont
				{M.}~\bibnamefont {Blondel}}, \bibinfo {author} {\bibfnamefont
				{P.}~\bibnamefont {Prettenhofer}}, \bibinfo {author} {\bibfnamefont
				{R.}~\bibnamefont {Weiss}}, \bibinfo {author} {\bibfnamefont
				{V.}~\bibnamefont {Dubourg}}, \bibinfo {author} {\bibfnamefont
				{J.}~\bibnamefont {Vanderplas}}, \bibinfo {author} {\bibfnamefont
				{A.}~\bibnamefont {Passos}}, \bibinfo {author} {\bibfnamefont
				{D.}~\bibnamefont {Cournapeau}}, \bibinfo {author} {\bibfnamefont
				{M.}~\bibnamefont {Brucher}}, \bibinfo {author} {\bibfnamefont
				{M.}~\bibnamefont {Perrot}},\ and\ \bibinfo {author} {\bibnamefont
				{{{\'E}}douard Duchesnay}},\ }\href
		{http://jmlr.org/papers/v12/pedregosa11a.html} {\bibfield  {journal}
			{\bibinfo  {journal} {Journal of Machine Learning Research}\ }\textbf
			{\bibinfo {volume} {12}},\ \bibinfo {pages} {2825} (\bibinfo {year}
			{2011})}\BibitemShut {NoStop}%
		\bibitem [{\citenamefont {Bewley}\ \emph {et~al.}(2018)\citenamefont {Bewley},
			\citenamefont {Ewings}, \citenamefont {Le}, \citenamefont {Perring},\ and\
			\citenamefont {Voneshen}}]{PyChop}%
		\BibitemOpen
		\bibfield  {author} {\bibinfo {author} {\bibfnamefont {R.~I.}\ \bibnamefont
				{Bewley}}, \bibinfo {author} {\bibfnamefont {R.~A.}\ \bibnamefont {Ewings}},
			\bibinfo {author} {\bibfnamefont {M.~D.}\ \bibnamefont {Le}}, \bibinfo
			{author} {\bibfnamefont {T.~G.}\ \bibnamefont {Perring}},\ and\ \bibinfo
			{author} {\bibfnamefont {D.~J.}\ \bibnamefont {Voneshen}},\ }\href@noop {}
		{\bibinfo {title} {Pychop}} (\bibinfo {year} {2018})\BibitemShut {NoStop}%
	\end{thebibliography}

\end{document}